\documentstyle[preprint,psfig,aps]{revtex}
\tightenlines
\def\lesap{\mbox{\hspace*{.15cm}\parbox[]{.4cm}{$\stackrel{\scriptstyle <}{\scriptstyle \sim}$}}}
\begin{document}
\title{Modelling $J/\psi$ production and absorption in a microscopic
nonequilibrium approach}
%\author{C.~Spieles$^1$\footnote{Supported by the Alexander v.~Humboldt Foundation}\footnote{Email: cspieles@lbl.gov },
%R.~Vogt$^{1,2}$, L.~Gerland$^3$, S.A.~Bass$^4{}^*$, M.~Bleicher$^3$, 
%H.~St\"ocker$^3$, W.~Greiner$^3$, L.~Frankfurt$^{3,5}$, M.~Strikman$^6$}

\author{C.~Spieles$^1$\footnote{Supported by the Alexander v.~Humboldt Foundation}\footnote{Email: cspieles@lbl.gov },
R.~Vogt$^{1,2}$, L.~Gerland$^3$, S.A.~Bass$^4{}^*$, M.~Bleicher$^3$, 
H.~St\"ocker$^3$, W.~Greiner$^3$}

\address{$^1$~Nuclear Science Division,
Lawrence Berkeley National Laboratory,
Berkeley, CA 94720, USA}
\address{$^2$~Physics Department,
University of California at Davis, 
Davis, CA 95616, USA}
\address{$^3$~Institut f\"ur
Theoretische Physik,  J.~W.~Goethe-Universit\"at,
D-60054 Frankfurt a.M., Germany}
\address{$^4$~Department of Physics, Duke University,
Durham, N.C. 27708-0305, USA}
%\address{$^5$~School of Physics and Astronomy, Tel Aviv University, 69978
%Ramat Aviv, Tel Aviv, Israel}
%\address{$^6$~Department of Physics, Pennsylvania State University,
%University Park, PA 16802, USA}
\date{\today}
\maketitle
\begin{abstract}
Charmonium production and absorption in heavy ion collisions is studied with
the Ultrarelativisitic Quantum Molecular Dynamics model. We compare the
scenario of universal and time independent color-octet dissociation cross 
sections with one of distinct color-singlet $J/\psi$, $\psi'$ and $\chi_c$ 
states, evolving from small, color transparent configurations to their 
asymptotic sizes. The measured $J/\psi$ production cross sections in $pA$ 
and $AB$ collisions at SPS energies are consistent with both --- 
purely hadronic --- scenarios. The predicted rapidity dependence of 
$J/\psi$ suppression can be used to discriminate between the two 
experimentally. The importance of interactions with secondary hadrons and the 
applicability of thermal reaction kinetics to $J/\psi$ absorption are 
investigated. We discuss the effect of nuclear stopping and the role of 
leading hadrons. The dependence of the $\psi'/J/\psi$ ratio on the model 
assumptions and the possible influence of refeeding processes is also studied. 
\end{abstract}

\pagebreak

\section{Introduction}
Experimental data from the CERN SPS \cite{baglin,gonin,abreu,romana} 
clearly show that the yield of $J/\psi$-mesons in $pA$ and $AB$ collisions
is suppressed when compared to hard QCD production. A substantial portion 
of these data can be described in simple models of nuclear absorption with 
certain assumptions about the dissociation cross sections, {\it e.g.}
\cite{gerschel88,gerschel92,gavin94,blaizot,gavin,wong,kharzeev,vogtplb98,capella}.
Recent, preliminary data on $J/\psi$ production in Pb+Pb collisions
\cite{gonin,abreu,romana} have generated an intense and ongoing
controversy regarding the necessity of a novel suppression mechanism, 
{\it e.g.} a quark-gluon plasma \cite{matsui},
to explain the data or whether
the suppression can be due to hadronic interactions.
(For theoretical reviews, see
Refs.~\cite{reviewvogt,reviewkharzeev,reviewmuller,bassqgp}.)
Charmonium production and absorption has been studied first by means of
semianalytical models of heavy ion collisions and/or thermal rate calculations
\cite{gerschel88,gerschel92,gavin94,blaizot,gavin,wong,kharzeev,vogtplb98,capella,matsui,gavin88,vogt88,neubauer89,martins,gerland,mueller} 
as well as, more recently, within
microscopic hadronic transport models \cite{cassing,cassing2,sorge,geiss,kahana,mylett}. 
Some current models suggest purely hadronic
explanations of the suppression while others claim that this is not possible.
Here we use different model scenarios to
identify crucial theoretical and experimental conditions which have to be met
before definite conclusions can be drawn.

In simple analytical models of $J/\psi$ suppression, based on a semiclassical 
Glauber approach, nuclear absorption is calculated by 
convoluting nuclear density profiles with an assumed cross
section for dissociation by nucleons. 
The model parameters can generally be tuned to fit 
all $pA$ data as well as all the $AB$ data except Pb+Pb 
\cite{gerschel92,gavin94,blaizot,gavin,wong,kharzeev,vogtplb98,capella} .
Absorption by produced hadrons, comovers, is typically also included 
in these models \cite{gavin94,gavin,wong,kharzeev,vogtplb98,capella}. 
The comover dynamics is treated in a simple, schematic fashion, generally
assuming that the comover density scales with the number of
nucleon participants with the exception of Ref.~\cite{capella}. 
It is not straightforward to distinguish between these results
because of the variation of model parameters such as the
comover dissociation cross sections, nuclear density profiles, 
formation and freeze-out times, as well as including feed-down to the
$J/\psi$ from more massive charmonium states. 
However, except for Ref.~\cite{capella}, all these studies claim that
hadronic interactions alone cannot account for the $J/\psi$
suppression in all $pA$ and $AB$ data.

Implicitly, geometrical models of nuclear absorption such as Glauber
scattering assume that the 
nuclei traverse each other without distortion 
or energy loss. This seems to be
appropriate for the description of the hard Drell-Yan \cite{Al90,baglinDY} or 
$c \bar c$ \cite{E769} production processes and some authors also treat
charmonium dissociation perturbatively \cite{kharzeev94}.
However, the dissociation of charmonium need not necessarily be
associated  with a large momentum transfer. Therefore, 
nuclear absorption may result from soft processes
which have been shown to dominate the charmonium dissociation at SPS energies
\cite{gerland}. 
If nuclear absorption is a soft process, the 
nucleons may scatter and lose energy before interacting with a charmonium 
state. Microscopic transport models, which approximate the soft hadron
dynamics of a heavy ion collision as a cascade of binary scatterings,
include this nontrivial baryon dynamics and are therefore adequate tools for
the simulation of charmonium absorption. Moreover, they treat the complete
space-time evolution of secondary hadrons in a consistent
fashion.

However, the treatment of hard production processes as $g g \rightarrow c\bar c$ or
$q \bar q \rightarrow \mu^+ \mu^-$ is not obvious 
in the microscopic simulation of a soft hadronic cascade.
The Drell-Yan production cross section in $pA$ interaction is 
known to scale linearly with $A$ \cite{Al90,baglinDY}: 
$\sigma_{\rm pA}= \sigma_{\rm 0} A \, $. Because of the apparent charmonium 
absorption  discussed below there is no direct experimental evidence for 
linear scaling in charmonium production, or the hard nature of this
process, respectively. However, we assume hard production of charmonium since
measurements of open charm production in $\pi+A$ reactions show 
no significant deviation from $A$ scaling \cite{E769}. It should be
mentioned, though, that the open charm data at low transverse momentum 
suffers from low statistics. We also note that, even though charmonium
production does not scale as $A$, the exponent, 0.9 \cite{Bad83,Al91}, is
closer to unity than for lighter hadrons at low $p_T$ and $x_F$ 
which grow approximately as $A^{2/3}$ \cite{Ba83}.

The near linear scaling of the $c \bar c$ production 
in $pA$ reactions is a consequence of requiring a single perturbative 
parton-parton interaction. To simulate this 
microscopically, soft interactions of nucleons 
and the associated energy loss must be neglected.
In Refs.~\cite{cassing,cassing2}, $c\bar c$ production was implemented 
into the hadronic transport code
as a possible subchannel of the total $NN$ cross section. 
This is certainly not consistent with the picture of
perturbative $c\bar c$ production. 
In Ref.~\cite{geiss,kahana,mylett}, on the 
other hand, the space-time production points are obtained from hard 
processes in each event before the hadronic background simulation. 
Charmonium dissociation is also treated differently in the various models. 
Some simulations, \cite{cassing2,geiss,kahana}, assume that charmonium precursor 
states can be immediately dissociated by nucleons 
with constant cross sections larger than $5$~mb. In Ref.~\cite{mylett} the charmonium cross
sections evolve from small, color transparent states
while Ref.~\cite{cassing} gives no definite formation time.
Only Refs.~\cite{kahana,mylett} include feeddown from $\psi'$ and $\chi_c$ 
mesons. 
Based on rather different model treatments of the charmonium dynamics and the 
interconnection of hard and soft processes, all these 
studies claim that conventional hadronic scenarios are consistent with the 
preliminary Pb+Pb data.

In this work, we simulate three different scenarios of charmonium absorption
within the same calculational framework and compare the results.
We apply the Ultrarelativistic Quantum Molecular Dynamics~model v1.1 \cite{bigpaper}
for all calculations presented here.

In the first scenario, charmonium states are assumed to be produced as 
color-octet states which can be dissociated by nucleons only \cite{kharzeev}. 
In the second scenario, nuclear absorption is treated in the same way as the
first.
However, secondary hadrons may dissociate the fully formed $J/\psi$, 
$\psi'$ and $\chi_c$ states after a finite formation time.
Finally, we investigate a scenario in which the charmonia are 
produced as distinct color-singlet states and evolve from small, color transparent
configurations to their asymptotic sizes \cite{gerland,mylett}.

In Sec.~\ref{model}, the model framework and the distinct scenarios will be
discussed in detail. The results of the simulations are then presented and
discussed in Sec.~\ref{results}. In Sec.~\ref{etdist}, we examine the 
consistency of transverse energy spectra measured by different experiments
and compare them with our calculated distributions.
In Sec.~\ref{spacetime}, some qualitative features of charmonium suppression
and the space-time evolution of the system are discussed.
Sec.~\ref{absystematic} compares the quantitative $AB$ systematics of $J/\psi$
suppression within our scenarios to the data. After a short
discussion of the basic differences between semianalytical and microscopic
descriptions of the comover dynamics in Sec.~\ref{semivsmic}, we turn to the 
$\psi'/J/\psi$ ratio and its sensitivity to different model assumptions
in Sec.~\ref{psip}. In Sec.~\ref{elabdep}, the
dependence of $J/\psi$ suppression on the projectile energy is studied.
The key role of rapidity in discriminating between the different 
scenarios $J/\psi$ suppression is identified in
Sec.~\ref{rapdep}. Finally, we investigate the time dependence of 
$J/\psi$-comover collision energies and dissociation rates in
Sec.~\ref{secondaries}. The applicability of thermal reaction kinetics is
also discussed. In Sec.~\ref{conclusion}, we summarize our results and draw
conclusions.

\section{The model}
\label{model}
\subsection{The general framework}
\label{model_general}
We simulate the production and the absorption of charmonium states in heavy
ion collisions within the Ultrarelativistic Quantum Molecular Dynamics model
\cite{bigpaper}. This model has been shown to successfully describe final
rapidity distributions of negatively charged hadrons and net-baryons in
$p+p$ and  nucleus-nucleus collisions \cite{bigpaper}.
While the model prescription for the charmonium rescattering stage is
varied in order to study different conceivable scenarios, the initialization
of the nuclei and the simulation of the hard $c\bar c$ production process 
are the same for all calculations presented in this work.
The large number of parameters in the phenomenological microscopic 
model \cite{bigpaper}, especially in its collision term, have not been 
altered in the present study. The default values of these parameters are 
fit to data from elementary hadronic and 
p+A reactions over a range of energies. All concepts and parameters related to 
the treatment of heavy quarkonia in the model are specified in the following.

We apply perturbative QCD to the production of charmonium states by 
simulating nucleus-nucleus collisions in the impulse approximation.
This is done microscopically by allowing
freely streaming projectile and target nucleons to 
scatter without deflection or energy loss.
The nuclear dependence of the parton distribution functions is neglected.
The resulting 
space-time distribution of charmonium production points is inserted into the
evolving hadronic environment calculated with UrQMD 
since the rare quarkonium
production processes are only small perturbations on the heavy ion 
collision\footnote{Since we consider only exclusive charmonium production, 
the QCD factorization theorem is inapplicable.}.
Our model is thus designed to account for partonic and hadronic aspects of
the charmonium dynamics. 

In UrQMD, the initialization of the projectile and target 
nuclei is based on the Fermi-gas ansatz. The centroids of the 
nucleons are then randomly distributed within a sphere of radius 
\begin{equation}
R(A)= r_0 \left[ \frac{1}{2} \left( A+\left[ A^{1/3}-1\right] ^3
\right) \right] ^{1/3} \, ,
\end{equation}
\cite{bigpaper}. The effective interaction radius is somewhat larger than 
$R(A)$ since the individual 
nucleons have a finite size. In $pA$ reactions, a smeared out surface region of
$\Delta R=1.1$~fm, extending beyond $R(A)$, results from the geometrical 
$NN$ cross section, $\sigma\approx 40$~mb.
The parameter $r_0=\left( \frac{3}{4\pi\rho_0}\right)^{\frac{1}{3}}=1.14$~fm 
since the nuclear matter ground state density
$\rho_0=0.16\;\rm fm^{-3}$ is used in UrQMD. The above described initialization 
renders a good agreement with measured charge density distributions in nuclei
\cite{konopka_bass_diss}.

We use this initialization procedure for the simulation of hard
processes via perturbative QCD and the soft hadronic cascade, respectively. 
Consequently, the average nuclear density profiles are the same. 
However, in each event, the initial positions of the 
nucleons are determined independently for the two simulations after the
impact parameter $b$ is chosen.
Thus, we insert the $c\bar c$ states into the hadronic environment according
to an averaged distribution of hard production processes rather than
linking the hard and soft simulations event by event.
We therfore avoid strong and probably unphysical spatial correlations
between hard and soft processes\footnote{If the hard and soft simulations are linked event by event,
the resulting absorption probabilities sensitively depend on technical
details of the implementation, {\it e.g.} whether the position
of a produced $J/\psi$ is chosen to be exactly the arithmetic mean of the 
positions of the two incident nucleons or randomly distributed between them.
This seems to be unreasonable and is due to the naive geometrical 
interpretation of the interaction cross section in the semiclassical cascade.}.

In the microscopic UrQMD transport model \cite{bigpaper},
many nucleons are excited to baryonic strings, {\it i.e.} their constituent
quarks and diquarks are separated and propagate independently as the
endpoints of color flux tubes. A significant fraction of the constituent quarks
in the incident nucleons may thus appear as the core of leading mesons and
are allowed to interact immediately after their production while
leading baryons containing the remaining diquark 
interact with a reduced cross section in the UrQMD model.
A sharp distinction between nuclear absorption and absorption by
secondaries is not supported in this picture of nucleon break-up.

\subsection{Distinct scenarios}
We now introduce the scenarios for quarkonium suppression applied in our
simulations.

\subsubsection{Scenario~I}
This scenario follows Ref.~\cite{kharzeev}.
Charmonium states are assumed to be produced as color-octet `preresonance'
states, $|c\bar c g\rangle$, which
exist until after their passage through the nuclei is completed. 
These preresonance states can be dissociated by nucleons with a constant 
and universal cross section. In the $NN$ center-of-mass frame, the 
$|c\bar c g\rangle$ states are produced at rest.
Nuclear stopping and a change of the absorption
cross section due to string excitation of incident nucleons is not considered.
Then, in the UrQMD simulation, all but
the charmonium-nucleon interactions have zero cross sections.
$J/\psi$ dissociation by secondary hadrons is not taken into
account, since QCD calculations with heavy quarks predict a strong damping close
to threshold \cite{kharzeev94}. 
There is one free parameter, the $|c\bar c g\rangle$--nucleon cross section.

\subsubsection{Scenario~II}
According to this scenario, nuclear absorption is due to interactions of 
color-octet states. Thus, absorption by nucleons is calculated as in
Scenario~I, and the $|c\bar c g\rangle$ dissociation cross section is the first free
parameter. Thus, we follow the arguments of \cite{kharzeev} and the
`preresonance' hypothesis, respectively, as far as the nuclear absorption is
concerned.

Then, in a full UrQMD simulation, the environment of secondary hadrons is
calculated. After their finite formation time 
(on average, $\tau_F\approx 1$ fm/c) resulting from the 
string fragmentation used in the UrQMD model, secondary 
hadrons may dissociate the $J/\psi$, $\psi'$ and $\chi_c$, which are
assumed to be fully formed states by $\tau_F$. 
In order to clearly separate the nuclear absorption of color-octet 
states from the delayed interactions of produced mesons with the asymptotic 
color-singlet states, we exclude the prompt interactions of leading secondary 
hadrons. The interaction of the leading quarks is effectively accounted for
in the calculation of the nuclear absorption because the possible 
break-up of the incident nucleons due to a string excitation is ignored.

To calculate charmonium dissociation by secondary hadrons, it is
necessary to distinguish between the distinct charmonium states in the 
simulation.
The charmonium states are distributed according to their assumed 
production probability 
times their decay probability to $J/\psi$'s.
Thus 40\% of the final states are $\chi_c$'s, 55\% are $J/\psi$'s, and 5\% are 
$\psi'$s
\cite{gavai}. According to the degeneracy of the angular momenta, 
1/3 of the $\chi_c$'s are $\chi_{c10}$ states and 2/3 are $\chi_{c11}$ states.
Their momenta are assigned according to the parametrization \cite{ramona},
\begin{equation}\label{eq_momdist}
E\frac{d\sigma}{dMdp^3}\sim(1-x_F)^{3.55} \exp(-p_T\, 2.08\rm \,GeV^{-1})\;
.
\end{equation}
Unlike Ref.~\cite{kharzeev}, we assume that the comover dissciation cross
sections are predominantly nonperturbative at SPS energies,
as shown for the dissociation by nucleons\cite{gerland}.
The nucleon absorption cross sections, $X(c\bar c)+N$, were
derived in the framework of a nonrelativistic quarkonium model 
\cite{gerland}, obtaining
$\sigma(J/\psi N)=3.6$~mb, $\sigma(\psi' N)=20$~mb,
$\sigma(\chi_{c10} N)=6.8$~mb, and
$\sigma(\chi_{c11} N)=15.9$~mb. These cross sections are not 
relevant for nuclear absorption in this scenario which is assumed to be 
exclusively due to interactions of the hypothetical color-octet states, as discussed above.
The charmonium-meson cross sections
($X(c \bar c) +\pi$, $X(c \bar c) +\rho$, {\it etc.})
are obtained by scaling the nucleon absorption cross sections by 
a common factor $F$ so that $\sigma(X(c\bar c) M)=F\,\sigma(X(c\bar c) N)$.
In the additive quark model, $F=2/3$.
However, in this scenario it remains a second tunable parameter.
Thus, additional suppression of the comover dissociation cross sections
due to the relatively lower $X(c\bar c)+M$ collision energies 
can effectively be accounted for by adjusting $F$.
All comover interactions above the respective dissociation
thresholds are assumed to break up the charmonium state.

Charmonium dissociation is thus calculated in two 
different simulation modes for nuclear and comover absorption, 
respectively, assuming factorization of nucleon and comover absorption. 
Separate
survival probabilities for dissociation by nucleons, $S_N$, and 
mesons, $S_M$, are computed. The total
survival probability is then $S_T=S_N S_M$.
There are similarities between this scenario and the two-stage cascade 
used in Ref.~\cite{kahana}.

\subsubsection{Scenario~III}
Here, the charmonium states are produced as distinct 
color-singlet states. Their
dissociation cross sections evolve linearly from small, color transparent
configurations \cite{gerland}. 
Charmonium absorption is considered as a
nonperturbative process, thus nuclear stopping is included in the
calculation of nuclear absorption. In particular, nucleons which are excited 
to strings  in a prior $NN$ collision have a reduced cross section
since they exist only as leading diquarks during their formation time.
Charmonium states can interact with incident nucleons, leading quarks and 
diquarks, and secondary hadrons (after their formation time).
The production probabilities of the individual charmonium states and their
momentum distributions are chosen as in Scenario~II.
The rescattering cross sections for $X(c\bar c)+B$, assuming $B\equiv N$,
are taken from Ref.~\cite{gerland} (see Scenario~II).
We now let $F=2/3$ according to the additive quark model.
All baryon and meson collisions above the respective dissociation
thresholds are assumed to break up the charmonium state.
The cross sections correspond to the geometrical transverse radii
$r_T^i=\sqrt{\frac{\sigma^i}{\pi}}$ of the
charmonium states.
We use $\sigma^i$  to estimate the respective
formation times $\tau_F^i$ of the charmonium states by choosing
$\tau_F^i=r_T^i/c$.
During these formation times
the cross sections increase linearly with $t$ \cite{gerland},
starting from $\sigma_0=0$ at $t=0$.
Since all the parameters have been fixed either by prior model calculations
\cite{gerland} or the dynamical model, UrQMD, itself, we do not adjust any
parameters in this scenario. 

%To avoid double counting, interactions of $c \bar c$ states with produced
%hadrons in $NN$ collisions are excluded in the model. 
%Absorption of $c \bar c$ states due to interactions with
%projectile and target nucleons is reduced by a factor $(A_P-1)/A_P$ and 
%$(A_T-1)/A_T$, respectively. This is done stochastically by rejecting
%individual dissociation processes with the probability $1/A_P$ and $1/A_T$,
%respectively. However, in $AB$
%collisions, all soft $NN$ collisions can contribute to the hadronic
%environment via production of secondaries which may dissociate 
%the $c \bar c$ states. The remaining error
%imposed by this concept --- inherent to all microscopic and analytical
%models of comover absorption --- is estimated to be very small.

\section{Results and discussion}
\label{results}
\subsection{$E_T$ distributions}
\label{etdist}
\begin{figure}[b]
\vspace*{\fill}
\centerline{\psfig{figure=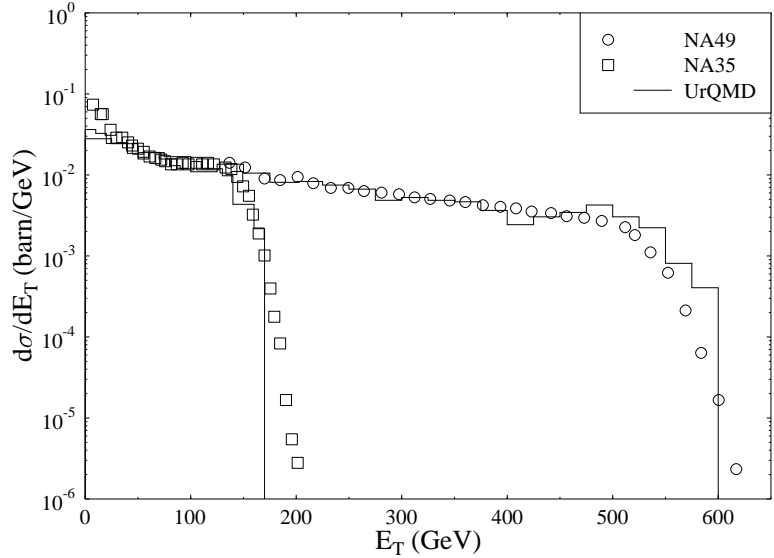,width=12cm}}
\caption{Differential cross section of the transverse energy produced in
Pb(160~GeV)+Pb and S(200~GeV)+Au collisions within the pseudorapidity range
$2.1<\eta<3.4$. The UrQMD results are compared with data
from \protect\cite{na49}.
\label{na49}}
\vspace*{\fill}
\end{figure}

The transverse energy distributions in different systems at 
SPS energies have been calculated with the UrQMD model \cite{bigpaper}. 
Figure~\ref{na49} shows the differential cross section of the transverse
energy produced in
Pb(160~GeV)+Pb collisions within the pseudorapidity range
$2.1<\eta<3.4$. The UrQMD result is compared with experimental data
from the NA49 Collaboration \cite{na49}. The overall agreement appears to be good 
although the
detailed shape of the spectrum at very central events, {\it i.e.} very high $E_T$,
is somewhat different. This difference may be due to statistical limitations
of the simulation since we have analyzed only 1000 events.
The disagreement between model and experiment at
very low $E_T$ can be due to an impact parameter cut at
$b=12.5$~fm in the calculations which leaves out part of the very
peripheral reactions.
The comparison between the S(200~GeV)+Au data from the NA35 Collaboration 
and the model calculation
within the same acceptance, $2.1<\eta<3.4$, shows an
agreement of similar quality, see Fig.~\ref{na49}.

The hard collision spectrum of neutral $E_T$ for
Pb(160~GeV)+Pb and S(200~GeV)+U reactions has also been calculated, 
based on the following concept:
First, the number of hard nucleon-nucleon collisions, which is proportional
to the yield of Drell-Yan muon pairs, is determined microscopically in the
simulation of freely streaming nucleons, as described in
Sec.~\ref{model_general}. The associated simulation of all soft processes
with the full hadronic cascade, initialized with the same impact parameter
$b$, then renders the information about the produced transverse energy.

\begin{figure}[b]
\vspace*{\fill}
\centerline{\psfig{figure=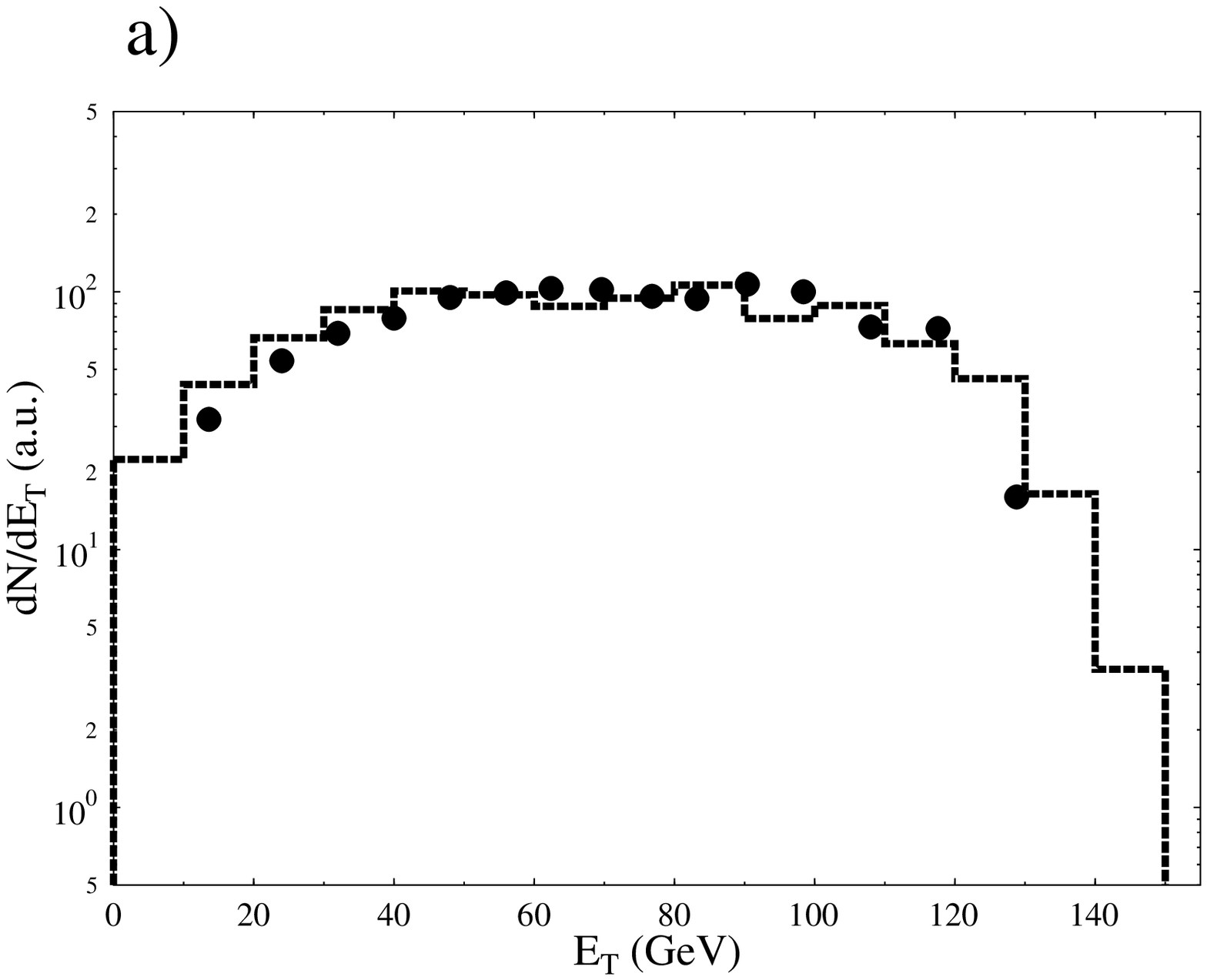,width=9cm}
\psfig{figure=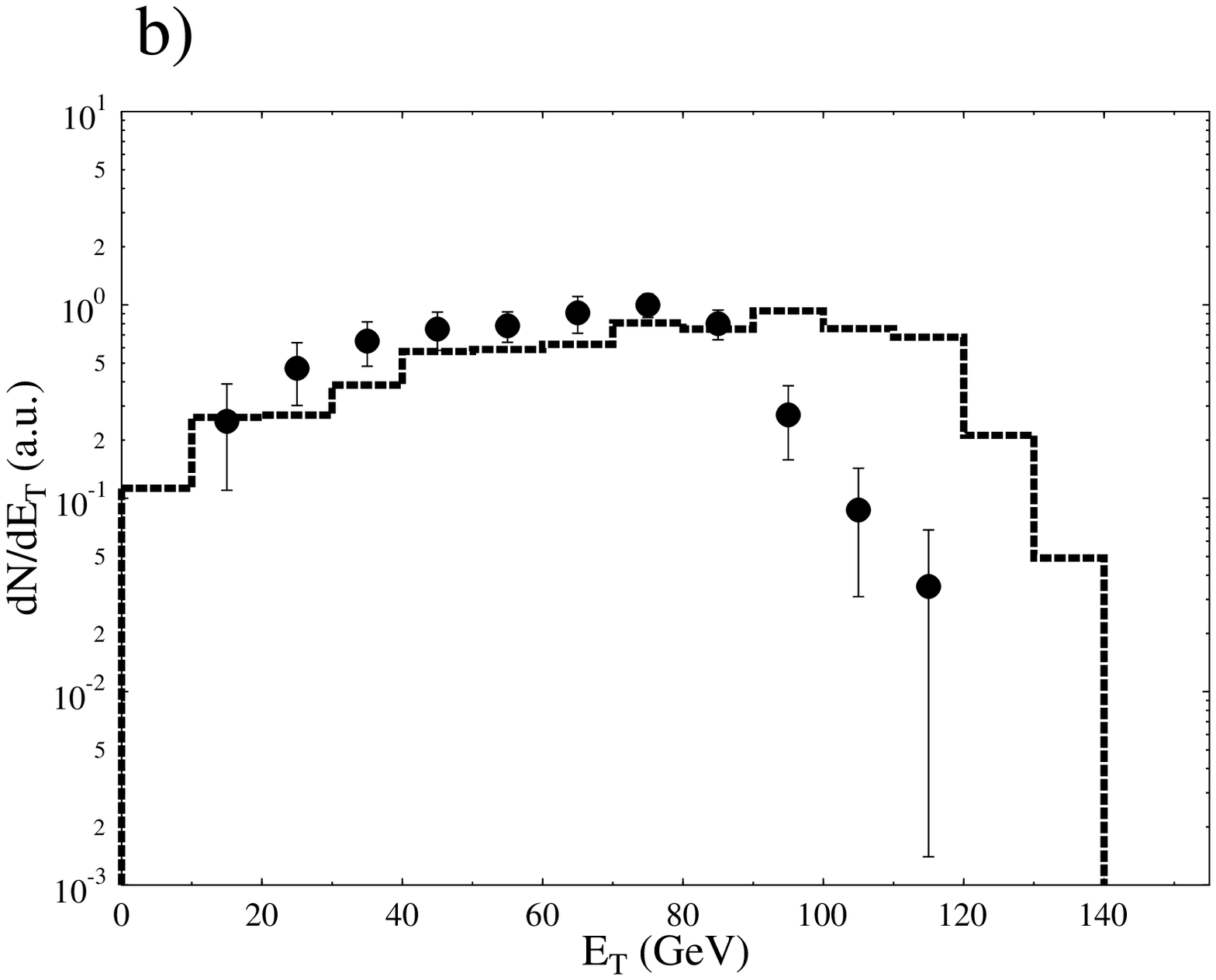,width=9cm}}
\caption{
a) Number of Drell-Yan muon pairs in Pb+Pb as a function of the produced neutral 
transverse energy within $1.1<\eta<2.3$ with arbitrary normalization.
The UrQMD result is shown with data from NA50
\protect\cite{ramello} where the
$E_T$-axis of the data is rescaled by $0.8$. The modification is motivated by the
recently published $E_T-E_{ZDC}$ contour plot \protect\cite{romana}
compared to the previously published analysis
\protect\cite{ramello,abreu}. 
b) Number of Drell-Yan muon pairs in S+U as a function of the produced
neutral transverse energy within $1.7<\eta<4.1$ with arbitrary
normalization. The UrQMD result and data from NA38
\protect\cite{borhani} are shown. 
\label{hardet}}
\vspace*{\fill}
\end{figure}

Figure~\ref{hardet}~(a) shows the number of Drell-Yan muon pairs 
($\propto$ number of collisions in the simulation of hard processes) in Pb+Pb 
interactions as a function of the produced neutral transverse energy within
$1.1<\eta<2.3$. The normalization is arbitrary. Experimental data from NA50
\cite{ramello} have been put on the plot as well. However, 
the abscissa has been rescaled since
the most recent data 
\cite{romana} indicate that the absolute $E_T$ values of the previous
publications on Pb+Pb collisions \cite{abreu,ramello} have shifted.
We are aware that the new analysis does not imply a simple
overall linear rescaling of the old data points. However, in order to 
reasonably compare the gross features of the
experimental dimuon $E_T$ spectrum 
with our model calculation, we have multiplied all $E_T$-values of the data in
the plot by 0.8.
This value was estimated by comparing the $E_T-E_{ZDC}$ contour plots of the
Tsukuba '97 \cite{ramello} and the Moriond '98 \cite{romana} proceedings.
The UrQMD simulation seems to be in rough agreement with the resulting
$dN/dE_T$ spectrum.
The  agreement between the model and experiment
is not unexpected since the production of the NA49 transverse energy had been shown
to be correctly described. The additional free streaming simulation
is very simple and represents a well understood model of hard scattering processes in nucleus-nucleus
collisions.

The agreement between model and experiment (NA38)
\cite{baglin,borhani} seems to become poor in the case
of the asymmetric S+U system, see Fig.~\ref{hardet}~(b). The UrQMD calculation
appears to significantly overestimate the neutral transverse energy within 
$1.7<\eta<4.1$
although the $E_T$ spectrum of the similar
S+Au system agrees with the NA35 data, as was shown in
Fig.~\ref{na49}.
Note that the use of a model for the simulation
of hard collisions and the weighting of the $E_T$ distribution with the
respective number of Drell-Yan dimuons
affects only the {\em shape} of the $E_T$ spectrum as compared
to a normal distribution of all events --- central events are
favored, since they correspond to a higher number of nucleon-nucleon
collisions. Thus, even if the correlation between transverse energy and
impact parameter was completely wrong in UrQMD, which seems unlikely given
the agreement in Fig.~\ref{na49},
this would merely lead to an incorrect weighting of the respective $E_T$ values.
In Fig.~\ref{hardet}~(b), however, one sees that the {\em maximally
achievable} neutral
transverse energy differs by about 25\%. 

In view of our simulation results, the NA38 neutral transverse energy
spectrum of S(200~GeV)+U seems to be inconsistent with the transverse energy
spectrum of S(200~GeV)+Au measured by NA35. On the other hand,
the (rescaled) NA50 Pb+Pb spectrum appears to be consistent with the NA49 
and the NA35 spectrum. Given the consistency of the microscopic simulation
with three different experimental measurements we assign the remaining
discrepancy in S+U to a possible problem in the determination of the  
NA38 $E_T$ scale. In order to compare our calculations of $J/\psi$ 
production as a function of
$E_T$ with NA38 data, we will therefore apply a scaling factor of
1.25 to the $E_T$ values reported by NA38 \cite{baglin,borhani}.

\subsection{Space-time picture of charmonium production and absorption}
\label{spacetime}
\begin{figure}[b]
%\vspace*{\fill}
\vspace*{-7.1cm}
\centerline{\hspace*{3cm}\psfig{figure=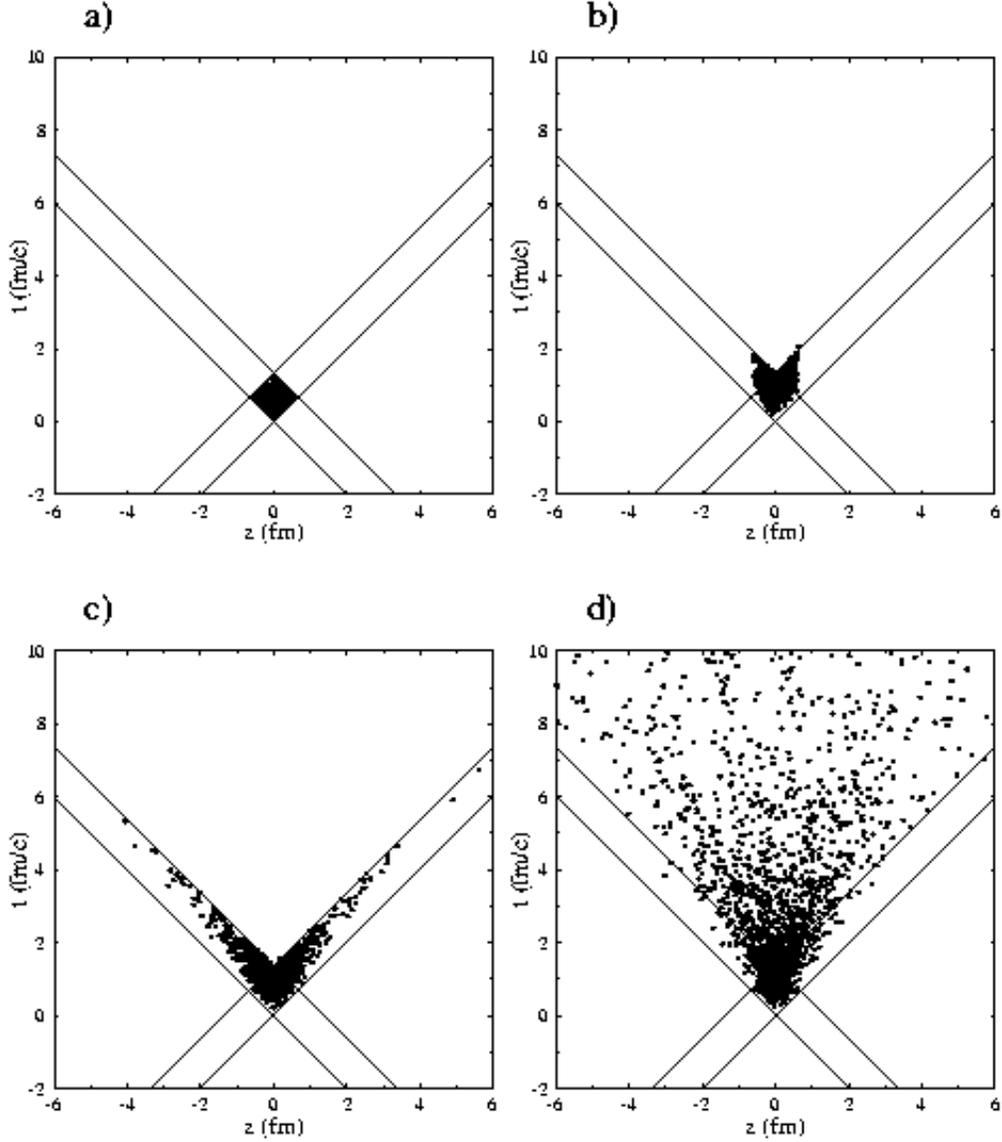,width=18cm}}
\vspace*{1cm}
\caption{Time evolution of a central Pb+Pb collision at $E_{\rm
lab}=160$~GeV according to the UrQMD simulation with time independent
absorption cross sections. The interaction
points of charmonium states in the $z$-$t$-plane are shown. 
a) The $c \bar c$ production points.
b) The $c \bar c$ dissociation
points due to interactions with nucleons, if the momentum distribution of 
the charmonium states and the nuclear stopping is neglected.
c) The momentum distribution of the charmonium states is now included
while the nuclear stopping is still neglected. d) The momentum distribution
of the charmonium states and nuclear stopping is taken into account.
\label{4zt}}
\vspace*{\fill}
\end{figure}

In this section, we illustrate how charmonium production and absorption
is incorporated into UrQMD.
Figs.~\ref{4zt}~(a)-(d) visualize the production and absorption points of
$c\bar c$ states in the $z$-$t$ plane for central Pb(160~GeV)+Pb collisions as
calculated with UrQMD. For this qualitative analysis, time independent 
baryon and meson absorption cross sections have been used. Their absolute 
value is irrelevant for this discussion. The impact of a time dependence of the cross sections on the 
dissociation probabilities will be discussed later in this section.

The straight lines indicate the light cone
trajectories of the geometrical boundaries of the nuclei. Hard production
processes occur only at times and positions where the nuclear densities
of projectile and target actually overlap in the course of the evolution.
Since the nuclei traverse each other without energy loss in the present 
simulation
of hard processes, the overlap region is determined by the light cone bands
in the plot. Figure~\ref{4zt}~(a) shows the $c\bar c$ production points.
They all lie in the nuclear overlap region. 
Figure~\ref{4zt}~(b) shows the points of $c\bar c$ dissociation by `hard
nucleons' in the $z$-$t$ plane, {\it i.e.} nuclear stopping is not included
in the simulation of nuclear absorption. Therefore, nuclear absorption is
constrained to the region covered by the light cone bands of the two nuclei,
however not to the actual overlap region of both bands.
The charmonium states are produced at rest. Therefore, the spatial
distribution of absorption points must coincide with the distribution
of production points, {\it i.e.} it is restricted to its longitudinal range 
$|z|<0.8$~fm. The temporal distribution of $c\bar c$ dissociation, however,
naturally exhibits some delay as compared to the production, since
absorption does not take place instantaneously.

Fig.~\ref{4zt}~(c) shows the points of $c\bar c$ dissociation by
nucleons as in Fig.~\ref{4zt}~(b), but
the charmonium states are now produced according to their assumed momentum
distribution given by Eq.~(\ref{eq_momdist}). Therefore, the spatial coordinates of production and
absorption do not coincide.
However, due to their high mass, few charmonium states actually acquire 
velocities high enough to propagate for very long and still be within 
reach of the nuclei which did not suffer energy and velocity loss in this calculation.

Fig.~\ref{4zt}~(d) shows the points of $c\bar c$ dissociation by `soft
nucleons' in the $z$-$t$ plane, {\it i.e.} nuclear stopping is included
in this simulation of nuclear absorption. Nuclear absorption can therefore
occur anywhere within the forward light cones of the two nuclei. Most of the
nuclear dissociation processes, however, take place at very early times,
$t\lesap 3$~fm/$c$.

\begin{figure}[b]
%\vspace*{\fill}
\vspace*{-7.2cm}
\centerline{\hspace*{3cm}\psfig{figure=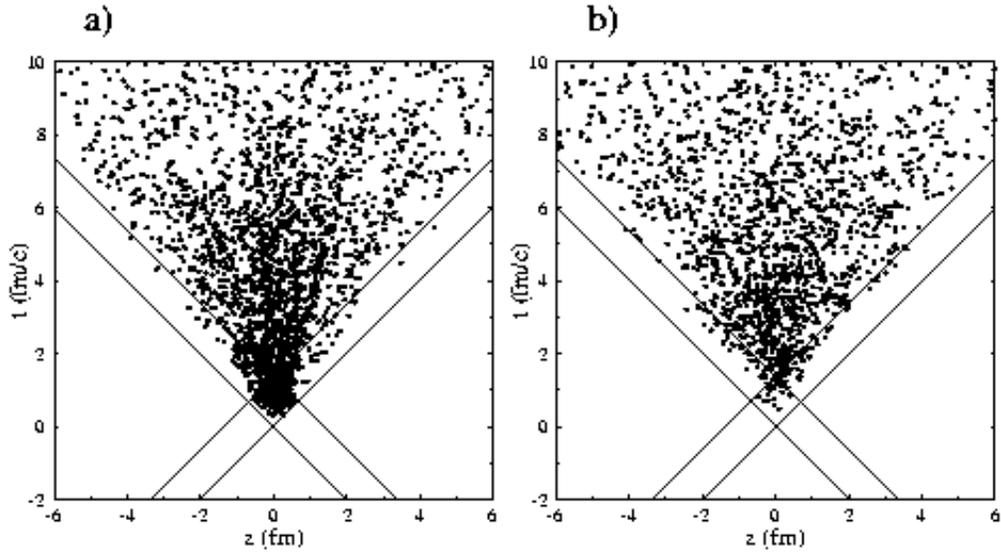,width=18cm}}
\vspace*{1cm}
\caption{Time evolution of a central Pb+Pb collision at $E_{\rm
lab}=160$~GeV according to the UrQMD simulation with time independent
absorption cross sections. The interaction
points of charmonium states in the $z$-$t$-plane are shown. 
a) The $c \bar c$ dissociation
points due to interactions with mesons including leading hadrons.
b) The $c \bar c$ dissociation
points due to interactions with mesons without leading hadrons.
\label{2zt}}
\vspace*{\fill}
\end{figure}

Fig.~\ref{2zt}~(a) shows the points of $c\bar c$ dissociation by 
mesons in the $z$-$t$ plane for central Pb(160~GeV)+Pb collisions. 
Note that mesons containing a constituent quark of an incident nucleon 
are included in this plot. These leading
hadrons can interact with other particles instantaneously, however, 
their cross section is reduced by a factor of two. Most of the absorption 
processes within the first
1.5~fm/$c$ are due to interactions with leading mesons.
To demonstrate this, Fig.~\ref{2zt}~(b) shows the points of $c\bar c$ 
dissociation by mesons in the $z$-$t$ plane, where only interactions of 
fully formed mesons are taken into account. Clearly, if leading mesons are
considered as comovers, they dominate the comover absorption probability.
Even a strong damping of the dissociation
cross section close to threshold \cite{kharzeev94} would not justify
neglecting these processes, since they occur well above threshold
(see Sec.~\ref{secondaries}).
On the other hand, nuclear stopping reduces the nuclear
absorption considerably. This is due to the fact that when a nucleon is
excited into a string, it loses one of its constituent quarks which
becomes part of a leading meson. 
Accordingly, the dissociation cross section of the baryon
containing the remaining leading diquark is rescaled to 2/3 of the nuclear
absorption cross section in the UrQMD simulation. 

\subsection{$AB$ systematic of $J/\psi$ suppression}
\label{absystematic}
Here we discuss the systematics of $J/\psi$ production and suppression as a
function of $AB$ in our three scenarios. We compare our results to the data
from several projectile-target combinations: 
$p$(450~GeV)+C, $p$(200~GeV)+Cu, $p$(200~GeV)+U, S(200~GeV)+U, 
and Pb(160~GeV)+Pb. The 450~GeV and 160~GeV
simulations are rescaled to $p_{lab}=200$~GeV with the parametrization of
Ref.~\cite{schuler}, as done by NA50 \cite{gonin,abreu}.

\subsubsection{Scenario~I}

\begin{figure}[b]
\vspace*{\fill}
\centerline{\psfig{figure=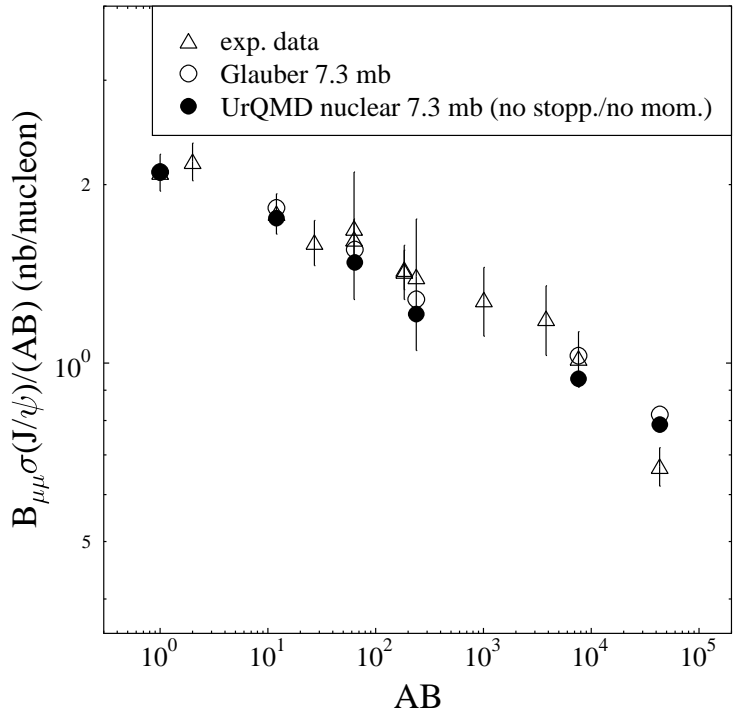,width=12cm}
}
\caption{$J/\psi$-production cross sections times dimuon branching ratio
in the kinematical domain $0<y_{cm}<1$ and $|\cos\theta_{CS}|<0.5$, 
and rescaled, if necessary, to $p_{\rm lab}=200$~GeV as a function of
$AB$. Open circles denote the semi-analytical Glauber calculations of
\protect\cite{kharzeev} while the results of the microscopic
UrQMD simulation without nuclear stopping and momentum distribution of the
$J/\psi$, Scenario~I, are shown as full circles. 
A universal and constant absorption
cross section of $\sigma=7.3$~mb is used for both.
The data (open triangles) are from \protect\cite{abreu}.
\label{khz}}
\vspace*{\fill}
\end{figure}

Figure~\ref{khz} shows a comparison between the semianalytical Glauber calculations
of Kharzeev {\it et al.} \cite{kharzeev} and UrQMD/Scenario~I using the same absorption
cross section, $\sigma^{\rm diss}=7.3$~mb.
The experimental data are also shown \cite{abreu}.
The $J/\psi$, $\psi'$
and $\chi_c$ are assumed to interact with nucleons while in $|c\bar c
g\rangle$ color-octet states. 

\begin{figure}[b]
\vspace*{\fill}
\centerline{\psfig{figure=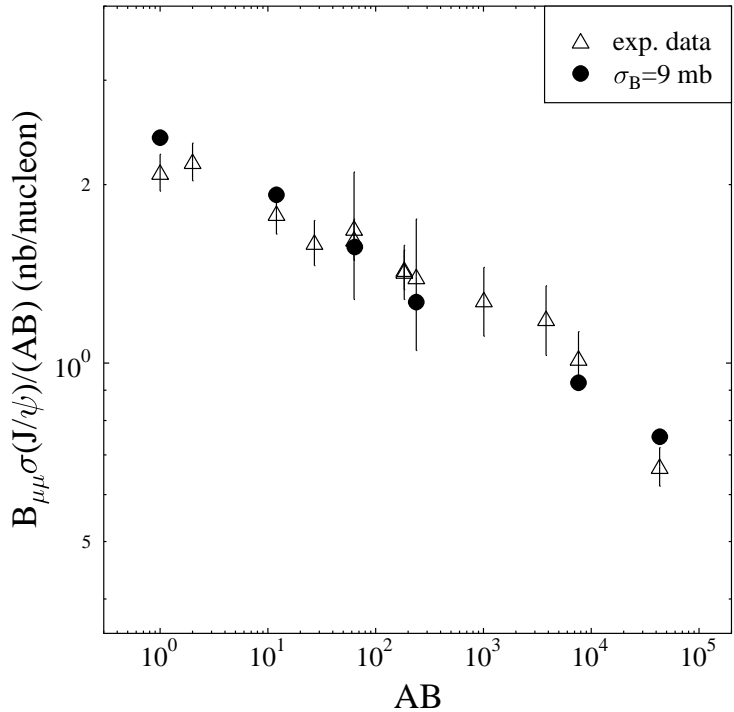,width=12cm}
}
\caption{$J/\psi$-production cross sections times dimuon branching ratio
in the kinematical domain $0<y_{cm}<1$ and $|\cos\theta_{CS}|<0.5$, 
and rescaled, if necessary, to $p_{\rm lab}=200$~GeV as a function of
$AB$. 
The results of the microscopic
UrQMD simulation without nuclear stopping and momentum distribution of the
$J/\psi$, Scenario~I, are shown as full circles. 
A universal and constant absorption
cross section of $\sigma=9$~mb is used. The $J/\psi$ production cross
section in p+p reactions (the absolute normalization) is arbitrarily chosen.
The data (open triangles) are from \protect\cite{abreu}.
\label{ver17}}
\vspace*{\fill}
\end{figure}

Charmonium dissociation by secondaries is assumed to
be negligible. As in the geometrical
picture, the $c\bar c$ states are produced at rest and the nuclei traverse
each other without energy loss. Nucleons do not suffer a reduction of their
cross section due to string excitations. The difference 
between the semianalytical model and the microscopic simulation according to
Scenario~I is 10\% in S+U
collisions ($S_{\rm Glauber}=0.49$ vs. $S_{\rm UrQMD}=0.45$). This is due to
the different nuclear density profiles:
In Ref.~\cite{kharzeev} a Woods-Saxon shape is used, whereas UrQMD
initializes hard spheres with an effectively smeared-out surface layer
due to the finite geometrical $NN$ cross sections, see Sec.~\ref{model_general}.
Note that the cross sections are derived by multiplying the respective 
survival probabilities with the experimental $p+p$ cross section
\cite{abreu}. In Ref.~\cite{kharzeev}, $p$+D reactions have been
used as reference value for determining the nuclear absorption cross
section. Surprisingly, the $p+$D point is above the $p+p$ point, measured at the same
energy. Note that Fig.~\ref{khz} is not our best fit
through the $pA$ and S+U data.
As in Ref.~\cite{kharzeev}, the present (analogous) calculation 
underpredicts the  $J/\psi$ absorption in Pb+Pb, while reasonable 
agreement for all other 
systems is achieved. However, a value of 9~mb for the color-octet 
absorption cross section yields an acceptable simultaneous agreement 
with all data if the $p+p$ cross section is left as a free parameter,
as shown in Fig.~\ref{ver17}.  
Note that our $p+p$ value differs from
the data by two standard deviations in this case.

%Fig.~\ref{stopp} shows the effect of nuclear stopping on the total 
%absorption. In other words, nuclear absorption is treated once as 
%hard process as in Scenario~I and once as a soft process. Constant 
%cross sections and  no $c\bar c$ momentum distributions are employed. 
%In $pA$ reactions, 
%`soft nucleons' dissociate $c\bar c$'s slightly more 
%effectively than `hard nucleons'. This seems to be reasonable because 
%the incoming nucleons are delayed due to multiple $NN$ collisions 
%and can therefore interact with a larger fraction of the produced $c \bar
%c$'s.

\subsubsection{Scenario~II}

\begin{figure}[b]
\vspace*{\fill}
\centerline{\psfig{figure=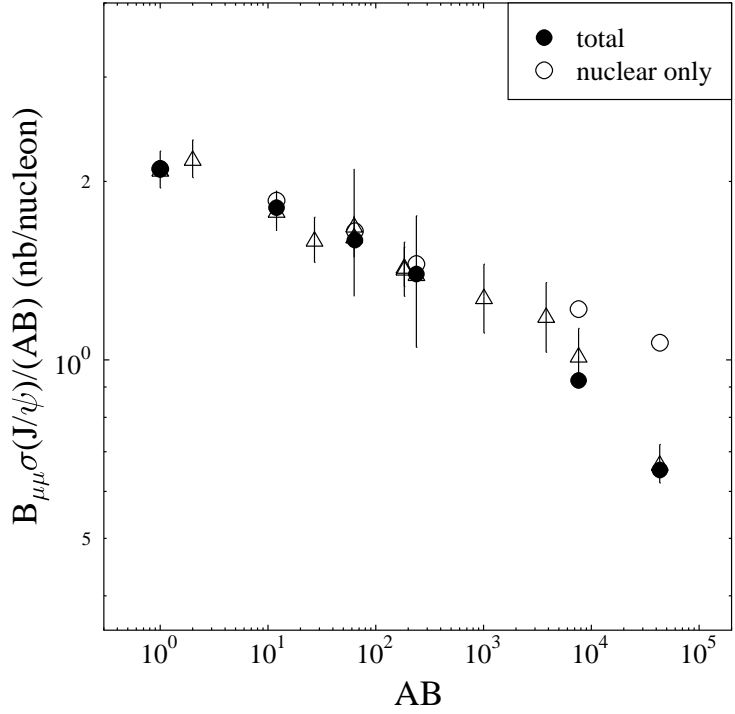,width=12cm}}
\caption{$J/\psi$-production cross sections times dimuon branching ratio
in the kinematical domain $0<y_{cm}<1$ and $|\cos\theta_{CS}|<0.5$, 
and rescaled, if necessary, to $p_{\rm lab}=200$~GeV as a function of
$AB$. Full circles denote the results according to Scenario~II.
Nuclear absorption is calculated  without stopping using a universal
dissociation cross section of $\sigma=4.8$~mb, {\it cf.}~Ref~\protect\cite{vogtplb98}. 
Comover absorption is
calculated in a full UrQMD simulation without leading hadrons. The
charmonium-meson cross sections ($X(c \bar c) +\pi$, 
$X(c \bar c) +\rho$, {\it etc.}) are reduced by a factor of 1/2 from the cross
sections for $X(c\bar c)+N$ from Ref.~\protect\cite{gerland}.
Open circles denote the production cross
sections if only nuclear absorption is considered.
The data (open triangles) are from \protect\cite{abreu}.
\label{sigab_ver15}}
\vspace*{\fill}
\end{figure}

Figure~\ref{sigab_ver15} shows a calculation of the $J/\psi$ systematics
within Scenario~II.
Recall that in order to avoid an overlap between
the interactions of the charmonia as color-octet and color-singlet 
states, leading mesons are not allowed to interact. The early stage of $c
\bar c$ absorption is thus described solely by Glauber-like nuclear
absorption using a universal color-octet dissociation cross section,
$\sigma(X(c\bar c) N)=4.8$~mb, {\it cf.}~\cite{gavin94}.
Interactions with secondary hadrons occur only after their formation time
--- on the average $\tau_F\approx 1$~fm/$c$ --- after the
charmonium states have fully formed. 
For the comover absorption cross sections
we take $F=0.5$
with the nucleon absorption cross sections from Ref.~\cite{gerland}.
The calculated $J/\psi$ production cross sections for the given parameter
values all lie within the experimental error bars. The measured cross
sections are therefore consistent with a purely hadronic picture of the
dynamics as described by Scenario~II. Note that similar model assumptions in a different
hadronic cascade model with a somewhat different choice of parameters have
lead to the same conclusion \cite{kahana}.
Semianalytical calculations, however, employing the same nuclear
dissociation cross section \cite{vogtplb98} fit the S+U data with
significantly smaller comover absorption cross sections. In contrast to the
microscopic calculation presented here, simultaneous agreement with the
Pb+Pb data cannot be achieved in that analysis, as will be discussed
in Sec.~\ref{micvsan}.

\subsubsection{Scenario~III}

\begin{figure}[b]
\vspace*{\fill}
\centerline{\psfig{figure=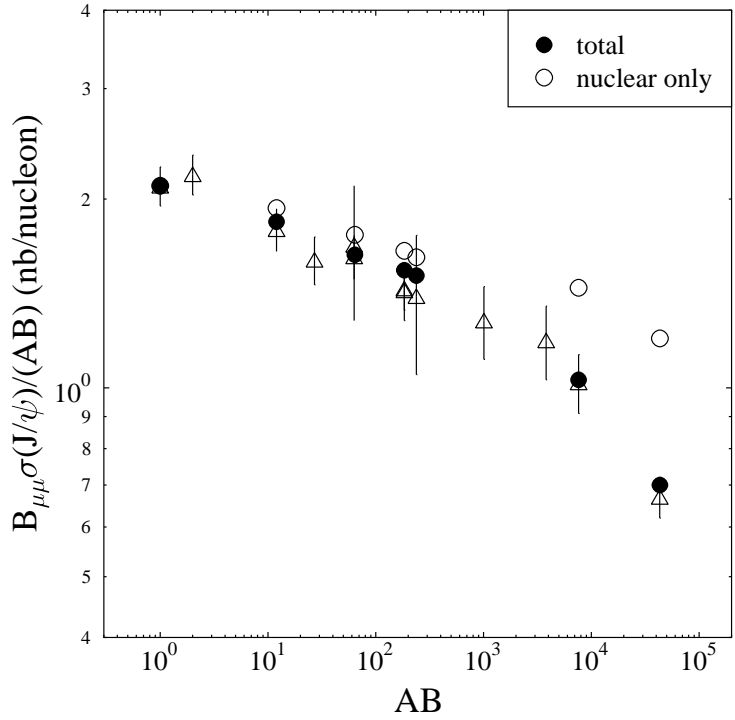,width=12cm}}
\caption{$J/\psi$-production cross sections times dimuon branching ratio
 in the
kinematical domain $0<y_{cm}<1$ and $|\cos\theta_{CS}|<0.5$, and rescaled,
if necessary, to $p_{\rm lab}=200$~GeV as a function of
$AB$. The UrQMD calculation with evolving dissociation cross
sections, Scenario~III, is shown.
Open circles denote the production cross
sections if only nuclear absorption is considered.
The data (open triangles) are from \protect\cite{abreu}.
\label{sigab}}
\vspace*{\fill}
\end{figure}

Figure~\ref{sigab} shows the UrQMD calculation assuming the charmonium
states are produced as color-singlet states and evolving from small, color
transparent configurations. Nuclear stopping and dissociation 
by leading  diquarks and quarks is taken into account. 
One observes a fair agreement of the
simulation with the $AB$ systematics of the data. 
The measured cross sections are therefore also consistent with a purely
hadronic system.
Note that no
parameters were tuned to fit the data.

\begin{figure}[b]
\vspace*{\fill}
\centerline{\psfig{figure=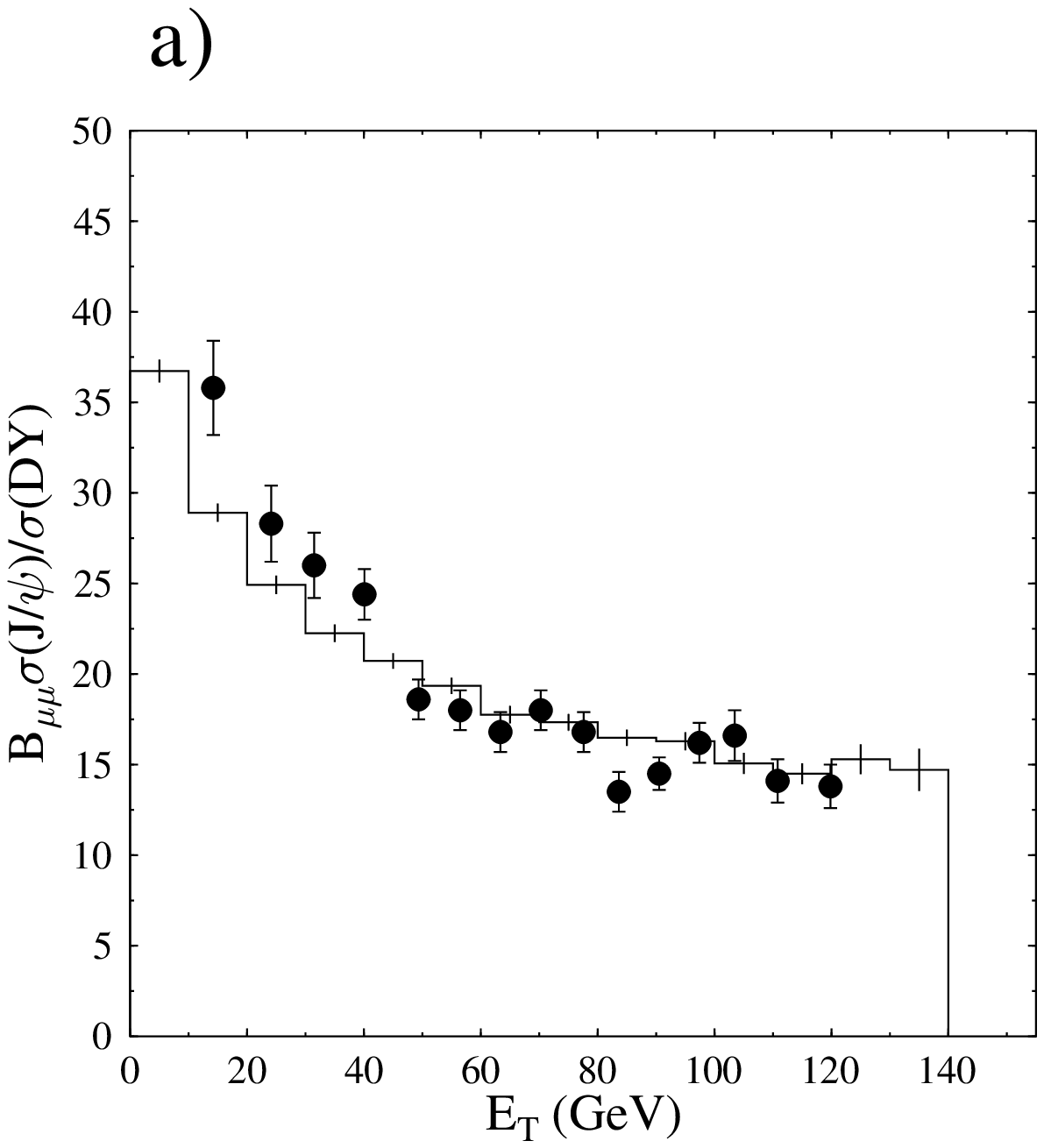,width=9cm}
\psfig{figure=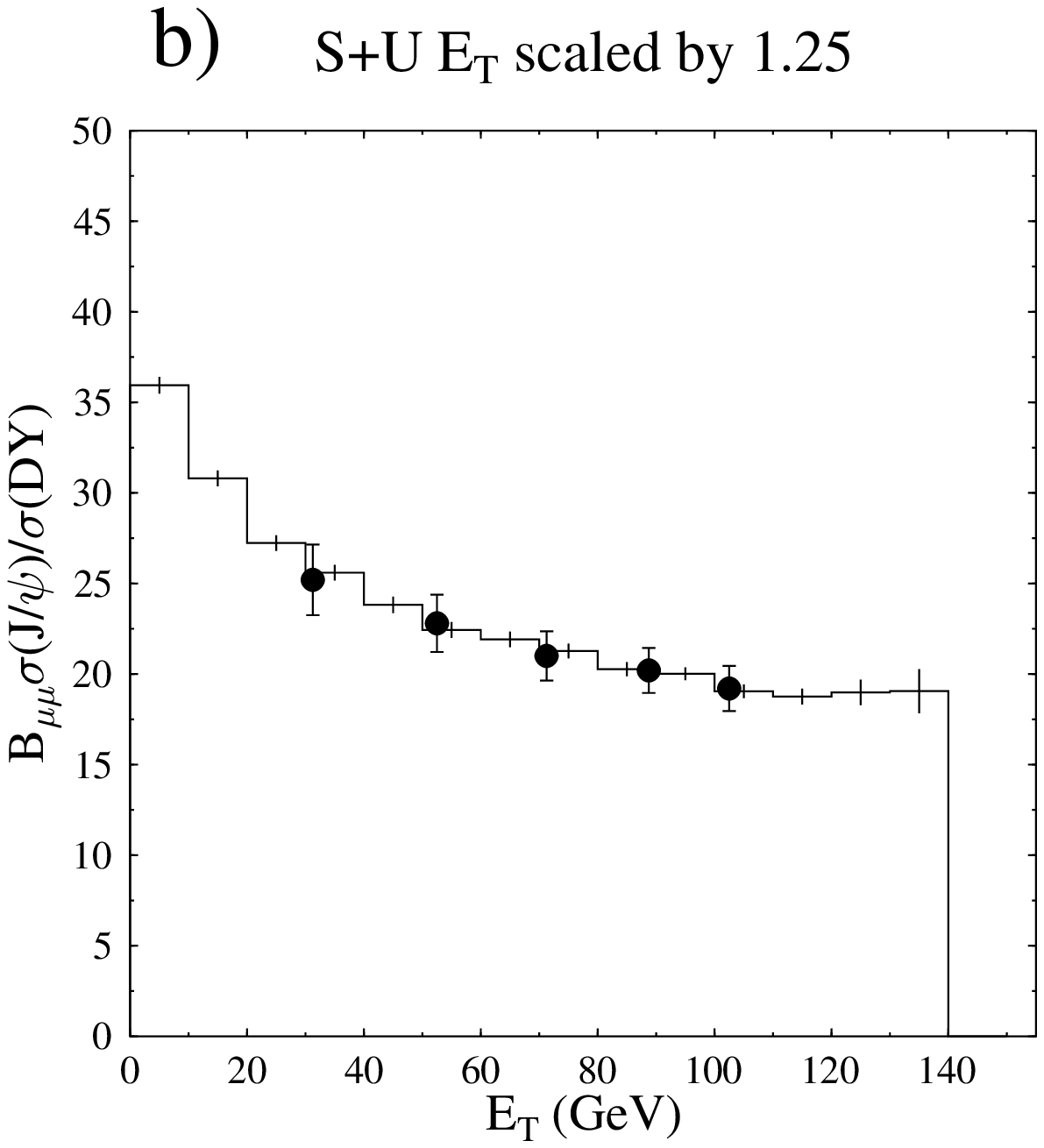,width=9cm}}
\caption{The ratio of $J/\psi$ to Drell-Yan production as a function of 
$E_T$ for Pb+Pb collisions at 158~GeV (a) and S+U at 200~GeV (b)
according to Scenario~III.
The experimental data are from Ref.~\protect\cite{romana} and
\protect\cite{gonin}, respectively.
The normalization factor $B\sigma_\psi/\sigma_{\mu\mu}=46$ in $pp$ 
interactions at 200~GeV is taken from Ref.~\protect\cite{kharzeev}. 
An additional factor of 1.25 
\protect\cite{reviewvogt} has been applied to the Pb+Pb calculation
in order to account for the lower energy (the
$J/\psi$ and the Drell-Yan cross section have a different energy
dependence).
The $E_T$ values of the experimental S+U data from
\protect\cite{gonin} are scaled by a factor 1.25 assuming that the UrQMD
calculation reflects the correct $E_T$ distribution, see Fig.~\ref{hardet}.
\label{psidyet}}
\vspace*{\fill}
\end{figure}

Figure~\ref{psidyet} shows the resulting $E_T$ distributions for the ratio of
$J/\psi$ to Drell-Yan production in Pb+Pb and S+U reactions. 
The gross features of the $E_T$ dependencies of the ratios are reasonably well
described by the model calculations. 
In particular, the calculated result is smooth in both cases, without abrupt 
discontinuities, in agreement with new high statistics data
comparing $J/\psi$ production to the minimum bias cross section
\cite{kluberg}. Apparent discontinuities in the preliminary data which
compared the more abundant $J/\psi$ to limited statistics 
Drell-Yan production had been
interpreted by some as a sudden onset of a deconfined phase
\cite{blaizot,kharzeevnardi,percolation}.

We want to point out that possible agreement or disagreement of the
simulation with the final data depends decisively on the determination of
the correct $E_T$ scale. When dealing with preliminary data, horizontal 
error bars indicating the systematic uncertainties in the $E_T$ scale of the 
experimental analysis would be helpful. The uncertainties of the
$E_T$ determination in our calculation are only statistical and lie within
the bin width of the histograms.

Note that the slight underestimation of the $J/\psi$ over Drell-Yan
production at low $E_T$ seen in Fig.~\ref{psidyet} can be related to an
overestimation of the Drell-Yan cross section at low $E_T$, visible in 
Fig.~\ref{hardet}~(a). Calculations in the context of the Dual Parton Model
also show have feature \cite{capella}.

\subsection{Semianalytical vs.\ microscopic descriptions}\label{micvsan}
\label{semivsmic}
It has been shown in the previous analysis that, within the microscopic
hadronic transport model UrQMD, comover interactions can explain
the observed $J/\psi$ suppression in S+U and Pb+Pb collisions.
The question then naturally arises why semianalytical calculations of charmonium 
production and absorption, based on the Glauber approach and the wounded
nucleon model, cannot also simultaneously describe these data
\cite{wong,vogtplb98}. Therefore, it is worthwhile to examine
some basic differences in the comover dynamics of
the two approaches. 

First, transverse expansion is neglected
in the semianalytical models \cite{wong,kharzeev,vogtplb98,capella} 
although it is a feature of the
true reaction dynamics, as established experimentally \cite{appels}. 
However, the full system including the transverse
dynamics \cite{bleicher} is 
treated explicitly in microscopic models like UrQMD.
It remains to be seen how the
$J/\psi$ survival probabilities are quantitatively affected by the
inclusion of a realistic transverse expansion in the semianalytic models.

Secondly, comover scenarios based on the wounded nucleon model assume a
linear scaling of the produced hadron rapidity densities with
the number of participants, $E{\rm d}^3N/{\rm d}p^3 \sim N_{\rm
part}^\alpha$ where $\alpha=1$. In contrast, a possible enhancement 
of the meson yield at midrapidity for nuclear collisions relative 
to N+N collisions has been observed \cite{baechler}. 
The $h^-$ yield per participant nucleon pair, extrapolated to full phase space,
is $4.0\pm0.2$ for Pb+Pb, compared to $3.6\pm0.2$ for S+S
\cite{na49preprint}. Measurements of the neutral pion production in Pb+Pb
collisions as a function of centrality show that the pion yield in the
interval $0.5<p_T<3$~GeV/$c$ scales like $N_{\rm part}^{1.3}$ \cite{wa98}.
Thus, experimental observations suggest deviations from
the simple assumptions of the wounded nucleon model at SPS energies.
This is not surprising since measured particle momentum correlation patterns
are in striking contrast to simple $p+p$ and $p+A$
extrapolations due to additional meson-meson and meson-baryon
interactions in nucleus-nucleus collisions \cite{proof}.

In UrQMD simulations of central Pb+Pb relative to central S+U
collisions the meson yield scales like $N_{\rm
part}^{1.25}$. This nonlinear scaling with the number of participants 
naturally explains at least a part of the ``additional'' suppression in the 
heavier system. Note that semianalytical
studies where the comover densities have been extracted from the Dual
Parton Model, which also exhibits a nonlinear scaling, produce 
simultaneous agreement with the S+U and Pb+Pb data \cite{capella}.

Finally, semianalytic models generally neglect the considerable 
rapidity asymmetry of hadron production  in asymmetric nuclear 
collisions \cite{na35dndy} while UrQMD simulations also show this feature of
the measured particle distributions. 
Even if the yields  
of produced hadrons over $4\pi$ scaled exactly with the number of 
participants, the
relative comover densities extracted from the wounded nucleon model 
would be incorrect in the finite experimental acceptance cut of NA50 , $0<y<1$.
According to the UrQMD simulation of central nuclear collisions,
$\frac{N^{h^-}({\rm Pb+Pb})}{N^{h^-}({\rm S+U})}|_{0<y<1}/\frac{N^{h^-}({\rm Pb+Pb})}{N^{h^-}({\rm
S+U})}|_{4\pi}=1.12$.

\subsection{The $\psi'$ to $J/\psi$ ratio}
\label{psip}
\begin{figure}[b]
\vspace*{\fill}
\centerline{\psfig{figure=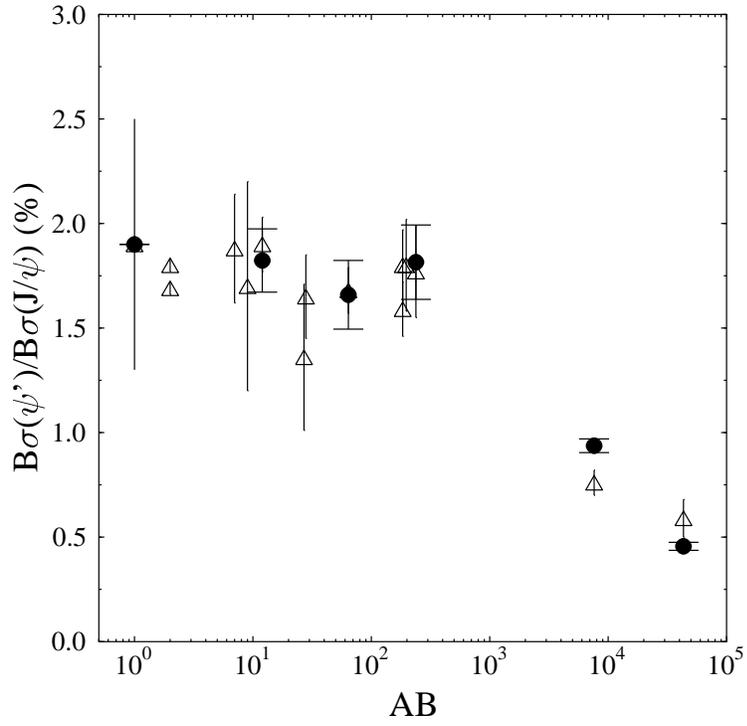,width=12cm}}
\caption{$\psi'/J/\psi$ ratio in $pA$, S+U and Pb+Pb interactions as function 
of $AB$ for universal nuclear dissociation cross sections and distinct
charmonium states interacting with comovers after their formation time, Scenario~II.
The data (open triangles) are from \protect\cite{gonin,lourenco}.
\label{ratab16}}
\vspace*{\fill}
\end{figure}

\begin{figure}[b]
\vspace*{\fill}
\centerline{\psfig{figure=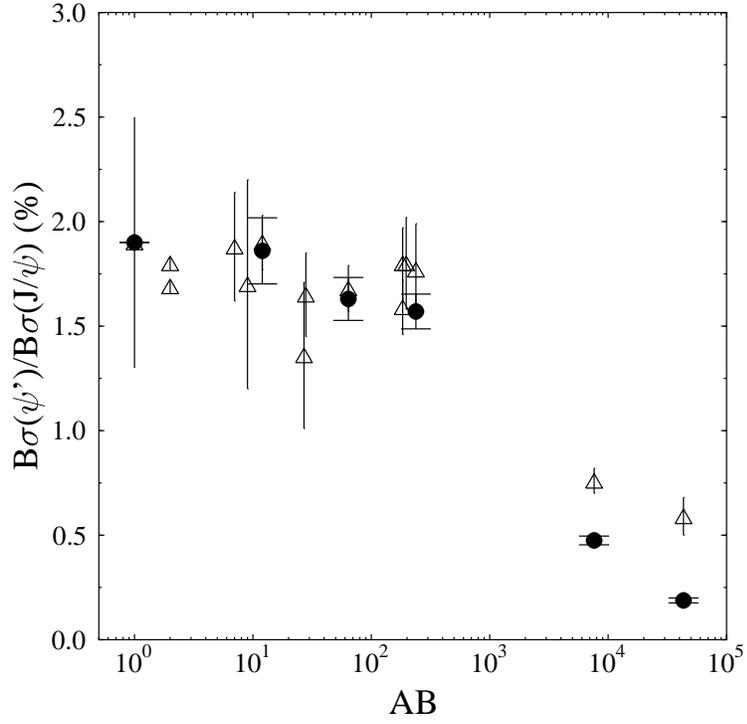,width=12cm}}
\caption{$\psi'/J/\psi$ ratio in $pA$, S+U and Pb+Pb interactions as function 
of $AB$ for evolving dissociation cross sections, Scenario~III.
The data (open triangles) are from \protect\cite{gonin,lourenco}.
\label{ratab}}
\vspace*{\fill}
\end{figure}

Figures~\ref{ratab16} and \ref{ratab} show 
the $\psi'/J/\psi$ ratio in $pA$ and $AB$ collisions 
as a function of $AB$. UrQMD calculations in Scenarios~II and III
are compared with the
experimental data \protect\cite{gonin,lourenco}.
The observed weak $A$ dependence of the ratio in $pA$ 
reactions is consistent with the hypothesis of 
the formation of a `preresonance' state, {\it e.g.}~\cite{kharzeev}. 
If the charmonium states are initially produced in the $|c\bar c g\rangle$
configuration and color neutralization takes sufficiently long,
the produced $J/\psi$'s and $\psi'$'s are equally suppressed by nuclear
absorption.
This is reflected in Fig.~\ref{ratab16}, showing the calculations within
Scenario~II. In $pA$ collisions, comover absorption is only a minor 
contribution to the total charmonium absorption. Consequently, the 
universal nuclear absorption cross section of the color-octet states 
results in a constant $\psi'/J/\psi$ ratio.

In Scenario~III, the absorption cross sections for the individual
charmonium states evolve linearly with time. Thus, the $\psi'$, 
$\chi_{c10}$ and $\chi_{c11}$ states have absorption cross
sections similar to that of the $J/\psi$ as long as $\tau^i < \tau_F^{J/\psi}$. 
Only for $\tau^i > \tau_F^{J/\psi}$ the
cross sections differ strongly, because then $\sigma(J/\psi N)$ has reached its
asymptotic value while the cross sections of the larger resonances still
increase. The differences in the calculated $J/\psi$ and $\psi'$ survival
probabilities for $pA$ reactions, as shown in Fig.~\ref{ratab}, are 
therefore mainly due to nuclear dissociation processes in the later stage 
of the reaction and interactions with secondaries.  
Even in $pA$ reactions comover interactions are not negligible in 
Scenario~III because leading mesons are taken into account.

Figures~\ref{ratab16} and \ref{ratab} also show the 
calculated $\psi'/J/\psi$ ratio in S+U and Pb+Pb reactions.
In Scenario~II, the $\psi'/J/\psi$ ratios in S+U and Pb+Pb reactions very roughly
agree with the experimental values.
In Scenario~III, dissociation by mesons leads to an even stronger decrease of 
the $\psi'/J/\psi$ ratio in S+U and Pb+Pb collisions than seen in the 
data \cite{gonin}. 
Although shortly after charmonium production the charmonium-comover 
dissociation cross sections  are similar for all states, 
the kinematic threshold leads to different dissociation probabilities.
%The influence on the $\psi'/J/\psi$ ratio of energy dependent cross 
%sections close to threshold still needs to be investigated.

The $\psi'/J/\psi$ ratios in nucleus-nucleus collisions seem to favor
Scenario~II as compared to Scenario~III, although neither
describes the data satisfactorily. However, the parameter space of the
models has certainly not yet been fully explored.
Moreover, for a complete understanding of the ratio, quantum interference
effects \cite{lonya} as well as refeeding processes such as 
$\pi J/\psi \rightarrow \psi' \pi$ \cite{sorge,chen} must be considered. 
As can be seen in Fig.~\ref{ratab}, the $\psi'$ survival probability
in S+U collisions in Scenario~III would need to 
be increased by more than 50\% in order to agree with the experimental observations.  
To test the effect of refeeding, we have implemented the process 
$\pi J/\psi \rightarrow \psi' \pi$ in UrQMD
with an energy independent cross section above the kinematic threshold, 
$E_{\rm th}=m_{\psi'}+m_\pi$. 
The refeeding cross section is assumed to have a time dependence like that
of the nuclear and comover absorption cross sections. In this case,
$\sigma_{\pi J/\psi \rightarrow \psi' \pi}=7$~mb would increase the
$\psi'/J/\psi$ ratio by 50\%.
This cross section seems unrealistically large because this exclusive
channel $\sigma$ is larger than the inclusive $\sigma(\pi\,J/\psi)=2.4$~mb
we have used elsewhere even though the mass threshold is larger than
$\pi\,J/\psi\rightarrow D\,\bar D$.
In order to further explore the effectiveness of possible $\psi'$ refeeding
processes, we have implemented the channel 
$M J/\psi \rightarrow \psi' M$, where $M$ stands for any meson species.
Then the refeeding cross section must be
$\sigma_{M J/\psi \rightarrow \psi' M}=3$~mb in order to increase the
$\psi'$ yield by 50\%. 

Finally, note that a change of the $\psi'$ dissociation cross section alone
evidently has a direct effect on the $\psi'/J/\psi$
ratio while the $J/\psi$ survival probability itself is hardly affected.
Therefore, in view of the theoretical uncertainties concerning charmonium 
dissociation cross sections and formation times it seems premature to 
draw definite conclusions
from the model comparison presented here.

\subsection{Pb+Pb at 160~GeV vs.\ 80~GeV}
\label{elabdep}
Since lowering the SPS beam energy has been discussed,
it is worthwhile to check if such a change could distinguish
between an evolving $c\bar c$ dissociation cross sections and a
constant cross section.
With a constant cross section,
nuclear absorption should be insensitive to the $\gamma$ factor, {\it i.e.}
the Lorentz contraction, of the nuclei because the reduced path length 
is canceled by the enhanced density. There is no cancelation, however, 
in the case of evolving cross sections. In the limit of very high beam
energies the nuclei become transparent for charmonium states despite the
large nuclear density, since the absorption cross section is essentially zero
immediately after production.
Comover absorption, on the other hand, should increase with energy in both
scenarios because the number of secondaries rises. However, the
average comover velocity should be also enhanced which increases
their average formation times in the center of mass frame
counteracting the effect of increasing density. 
Thus for very high energies, disregarding leading hadrons, 
absorption by comovers must eventually become negligible due to the
large formation times.
In any case, the comover density and its time evolution 
varies with energy and can lead to different characteristic behaviours within
the two scenarios. 

We have compared the calculations with the standard choice of 
parameters for the calculation with constant cross sections, Scenario~II,
and the evolving cross sections, Scenario~III. 
The resulting  survival probabilities for minimum bias Pb+Pb collisions 
at $E_{\rm lab}=160$~GeV and 80~GeV are presented in Table~\ref{table}.
In fact, the calculated energy dependence of the nuclear absorption within 
the two scenarios behaves as anticipated. Dissociation by nucleons stays
constant in Scenario~II while it decreases with increasing beam energy
in Scenario~III.
The numerical results show
that the comover absorption gets slightly stronger with increasing energy
both for constant and evolving dissociation cross sections.
The combined effects of absorption by baryons and mesons lead to a
decrease of the $J/\psi$ survival probability for 80~GeV as compared to
160~GeV in the case of evolving cross sections and an increase for constant
cross sections. However, the relative change of the survival probability 
is only on the order of a few percent in both cases. Thus the experimental 
distinction between the two scenarios with the suggested change of beam 
energy is unlikely.

\subsection{Rapidity dependence of $J/\psi$ suppression}
\label{rapdep}
Another means by which the charmonium production mechanism might be
determined is the study of the survival probability as a function of
rapidity. We investigate $pA$ interactions because absorption effects are
limited to only one nucleus and comover suppression is significantly reduced
in $pA$ relative to $AB$ collisions. A constant absorption cross section
should be independent of rapidity while an evolving cross section should
produce more absorption close to the target since the asymptotic state is
more likely to interact inside the target.

\begin{figure}[b]
\vspace*{\fill}
\centerline{\psfig{figure=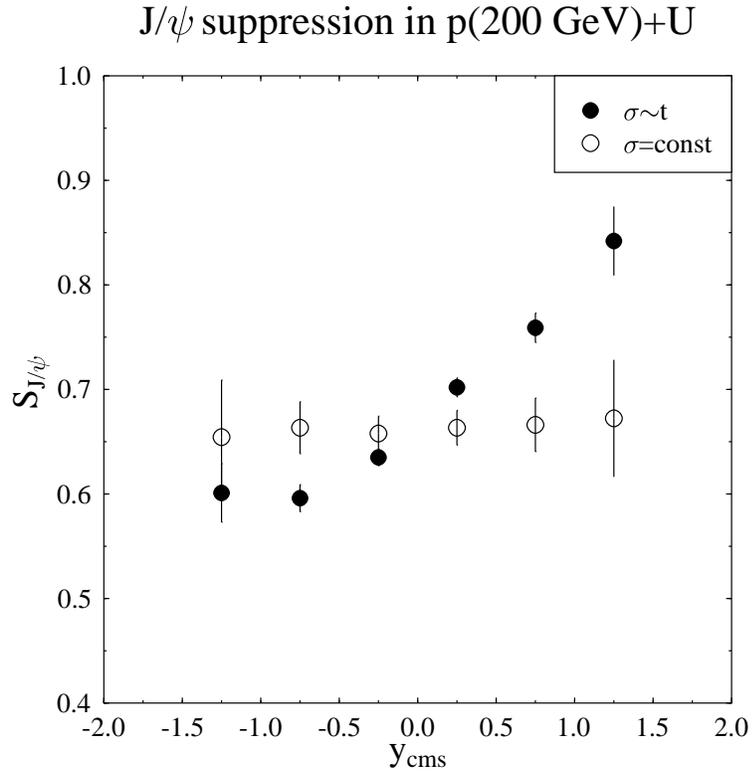,width=12cm}}
\caption{$J/\psi$-survival probability in p(200~GeV)+U reactions as a
function of rapidity.
Calculations with Scenario~II (open circles) and Scenario~III
(full circles) are shown.
\label{rapidity}}
\vspace*{\fill}
\end{figure}

Figure~\ref{rapidity} shows the rapidity dependence of the $J/\psi$
survival probability in p+U reactions at $E_{\rm lab}=200$~GeV. The
predictions of Scenario~II and Scenario~III are compared. The charmonium 
absorption is dominated by nuclear absorption in this system although
comover interactions are not completely negligible, in particular in
Scenario~III, see Figs.~\ref{sigab_ver15} and \ref{sigab}. 
The scenario of a constant color-octet dissociation cross section 
leads to a constant rapidity dependence of the $J/\psi$ survival probability
because the effective path length of charmonium
`preresonance' states does not depend on their longitudinal velocity as long
as $y_{X(c\bar c)}>y_{\rm target}$. The assumption of evolving 
color-singlet states leads to an increasing survival
probability with increasing rapidity. At forward rapidities, the larger
$\gamma$ factor of the charmonium states leads to an increased formation time
in the $NN$ center of mass frame. The evolution of the dissociation cross
sections is therefore delayed. The charmonium states 
suffer effectively less absorption during the passage through the target nucleus.
The reverse effect is limited by the asymptotic values of the absorption
cross sections. Therefore, the $J/\psi$ survival probability saturates at
negative rapidities. 

The experimental examination of the rapidity dependence in $pA$ reactions
is a very promising test of the production mechanism
because of the very distinct behaviour of the $J/\psi$ survival probability.
It would be particularly interesting to carry out this experiment at about the
same energy as the nuclear collisions studied by NA38 and NA50. 
Detailed analyses of existing experimental data on the
$x_F$ dependence of $J/\psi$ production in $pA$ reactions 
\cite{Bad83,Al91} should also be used to test the 
validity of different production mechanisms as done in Ref.~\cite{ramona_xf}.
There, it is also shown that the shape of the $x_F$ dependence 
changes with energy. It is conceivable that 
a realistic scenario of charmonium production and absorption involves a
superposition of color-singlet and color-octet contributions \cite{ramona_xf}, 
however, the relative importance may well be energy dependent.

\subsection{A closer look into interactions with secondaries}
\label{secondaries}

\begin{figure}[b]
\vspace*{\fill}
\centerline{\psfig{figure=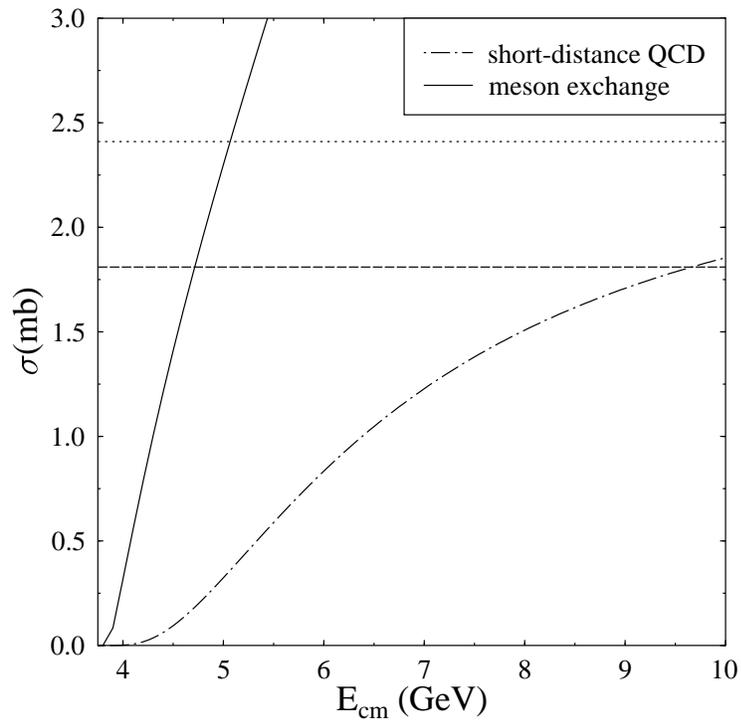,width=12cm}}
\caption{$J/\psi\,\pi$ dissociation cross section as calculated in 
short-distance QCD according to \protect\cite{kharzeev94} in comparison with 
a calculation with a meson exchange model \protect\cite{mueller}. Also shown
are the values of the $J/\psi$-meson absorption cross section 
chosen for the model calculations with evolving cross sections, $\sigma=2.4$~mb (Scenario~III), 
and for the calculations with constant cross sections, color-octet nuclear
absorption and no leading mesons, $\sigma=1.8$~mb (Scenario~II).
\label{xsections}}
\vspace*{\fill}
\end{figure}

From the results presented in Sec.~\ref{absystematic} we conclude that the
observed $J/\psi$ production cross sections in $pA$ and $AB$ collisions can
be explained in purely hadronic scenarios provided that
secondary particles can effectively dissociate charmonium. In
Scenarios~II and III we have employed universal and energy
independent\footnote{Above the respective kinematic dissociation threshold,
{\it i.e.} $\sigma_{\rm diss}(\sqrt{s})=\theta(\sqrt{s}-E_{\rm
th})\sigma_0 \,$ .}
comover absorption cross sections for the individual $c\bar c$ meson states,
see Fig.~\ref{xsections}. Given the $X(c\bar c)\,N$ cross sections
\cite{gerland}, the size of the comover cross sections --- with a reduction
of $F=1/2$ and 2/3, respectively --- are presumably of the right order, if
similar collision energies are considered.
However, calculations within the framework of short-distance QCD
\cite{kharzeev94} show that $J/\psi$ dissociation by hard gluons
is strongly damped close to the kinematic threshold.
Such a calculation is shown in Fig.~\ref{xsections}
for the process $J/\psi\,\pi$. This would significantly increase the 
survival probability. However, the comover absorption 
would not be completely suppressed because of the feeding from the less bound 
$\chi_c$ and $\psi'$ states. For example, the survival probability due to
interactions with comovers in minimium bias Pb(160~GeV)+Pb collisions in
Scenario~II increases from $S_M=0.61$ to $S_M=0.69$ if the $J/\psi$ itself
cannot be dissociated by mesons. As a result, the total survival probability
in this system would increase from $S_{tot}=0.31$ to $S_{tot}=0.35$.

Calculations of $J/\psi\,\pi$ and $J/\psi\,\rho$ dissociation cross sections 
in the framework of a meson exchange model also show a suppression at low
energies even for the exothermic $J/\psi\,\rho\rightarrow D \bar D$ process
\cite{mueller}. The cross section for the dissociation by pions according to
these calculations is shown in Fig.~\ref{xsections}. Note that the inclusion
of a form factor would further decrease the cross sections \cite{mueller}. 

From phase space arguments alone, {\it i.e.} assuming an energy independent
matrix element, one would infer an {\it enhancement} for low relative velocities
in the incoming channel if the dissociation process is exothermic. 
Calculations of cross sections involving heavier 
mesons in the incoming channels within meson exchange models would render
valuable information in order to estimate comover absorption probabilities. 
Also, a deeper understanding of the reactions involving the $\chi_c$ and 
$\psi'$ states is needed.

In a chemically equilibrated environment at temperatures less than $300$~MeV, 
the consideration of interactions with $\pi$'s and $\rho$'s might be
sufficient to quantitatively estimate the total comover absorption. 
In any kind of thermal hadronic scenario, higher
resonances are negligible due to the low density, $\rho_H\sim\exp{(-m_H/T)}$.
However, Table~\ref{table2} shows the relative importance of the mesons which
contribute most to $J/\psi$ dissociation according to the nonequilibrium 
UrQMD calculation. Scatterings of the $J/\psi$'s with $\pi$'s and $\rho$'s are
indeed the dominant dissociation processes. However, together they are
responsible for only about 40\% of the total comover absorption.
The composition is similar in Scenarios~II and III.
Twenty channels are left out of Table~\ref{table2}, each of them
contributing $\approx 1$\% or less. However, they account for
about 15\% of the total absorption.

The string fragmentation scheme employed in UrQMD \cite{bigpaper} determines 
the relative abundancies of hadron states in the nonequilibrium simulation. 
The choice of parameters is far from unique 
since the model can be compared to data only for experimentally identifiable
hadrons. 
In UrQMD, all states of eight different meson multiplets 
($J^{PC}=0^{-+},0^{++},1^{--},1^{+-},1^{++},2^{++}$ plus two excited
$1^{--}$ multiplets) may be populated in a string fragmentation process.
The probability to form a meson from one of these multiplets is chosen to be
proportional to the spin degeneracy and inversely proportional to the
average mass of the multiplet \cite{bigpaper}.

\begin{figure}[b]
\vspace*{\fill}
\centerline{\psfig{figure=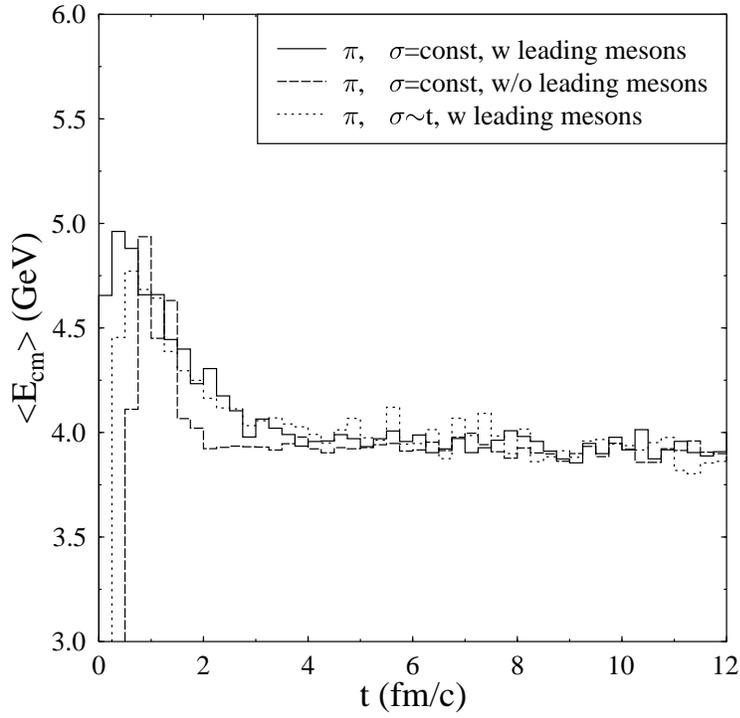,width=12cm}}
\caption{Average $J/\psi\, \pi$ collision energies 
in central Pb(160~GeV)+Pb reactions as a function of time for different
comover scenarios. We show the distributions for constant dissociation
cross sections including leading meson interactions (full line), excluding
leading mesons, Scenario~II (dashed line) and for evolving cross 
sections with leading mesons, Scenario~III (dotted line).
\label{ecmt}}
\vspace*{\fill}
\end{figure}

In Fig.~\ref{ecmt}, the average $J/\psi\, \pi$ collision energies as
functions of time are shown for three different scenarios: constant cross
sections including interactions of leading mesons
({\it cf.}~Ref.\cite{mycollspec}), constant cross sections excluding leading
mesons (Scenario~II), and evolving cross sections
sections including interactions of leading mesons (Scenario~III).
We studied central Pb(160~GeV)+Pb collisions.
All model simulations yield high average collision energies of
$\langle E\rangle\approx 4.8$~GeV at $t\approx 1$~fm/$c$. 
As stated in Ref.~\cite{mycollspec},
these collision energies are not compatible with an equilibrated hadron 
gas scenario since they correspond to a 
meson gas with a temperature of $T\approx 750$~MeV. 
The collision energies and the scattering rates at
later times, $t>2$~fm/$c$, are both consistent with temperatures of 
$T=140\pm20$~MeV \cite{mycollspec}, thus allowing for the application of 
a thermal reaction kinetics model.

\begin{figure}[b]
\vspace*{\fill}
\centerline{\psfig{figure=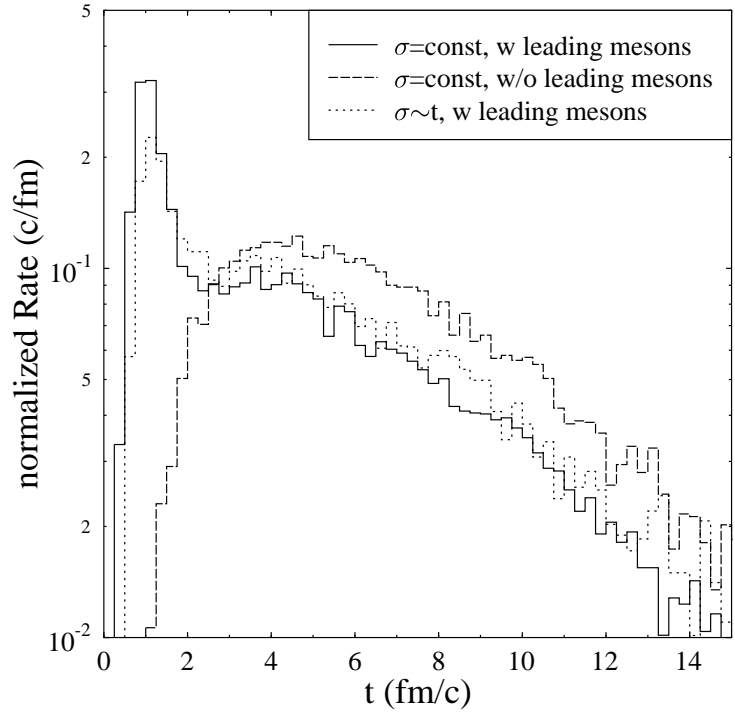,width=12cm}}
\caption{$J/\psi$-meson collision rates, normalized to one dissociation, 
in central Pb(160~GeV)+Pb reactions as a function of time.
We show the results for constant dissociation cross sections including
leading meson interactions (full line), excluding leading mesons,
Scenario~II (dashed line) and for evolving cross sections with leading mesons,
Scenario~III (dotted line). 
\label{rates}}
\vspace*{\fill}
\end{figure}

An important difference between the three scenarios presented in
Fig.~\ref{ecmt} is shown  in Fig.~\ref{rates} which determines the impact
of the highly nonthermal secondaries at early times on the total comover
absorption. 
The scattering rate as a function of time, normalized to
one $J/\psi$-meson collision, is depicted. It can be seen that  
at early times $t<2$~fm/$c$,
dissociation by comovers is
almost exclusively due to interactions of leading mesons. Moreover, if
leading meson interactions are taken into account, they clearly dominate the
total comover absorption. This is true for the scenario with constant as
well as with evolving cross sections. Note that the dissociation cross
section of a leading meson is only 1/2 of that of a fully formed meson.
As can be seen from Figs.~\ref{ecmt} and \ref{rates}, the applicability 
of thermal momentum distributions 
for the comover-charmonium interactions in heavy ion collisions depends heavily on the underlying model
assumptions. A thermal comover model seems to be a reasonable approximation
in Scenario~II
while it cannot be justified in Scenario~III where leading mesons are
included.

\begin{figure}[b]
\vspace*{\fill}
\centerline{\psfig{figure=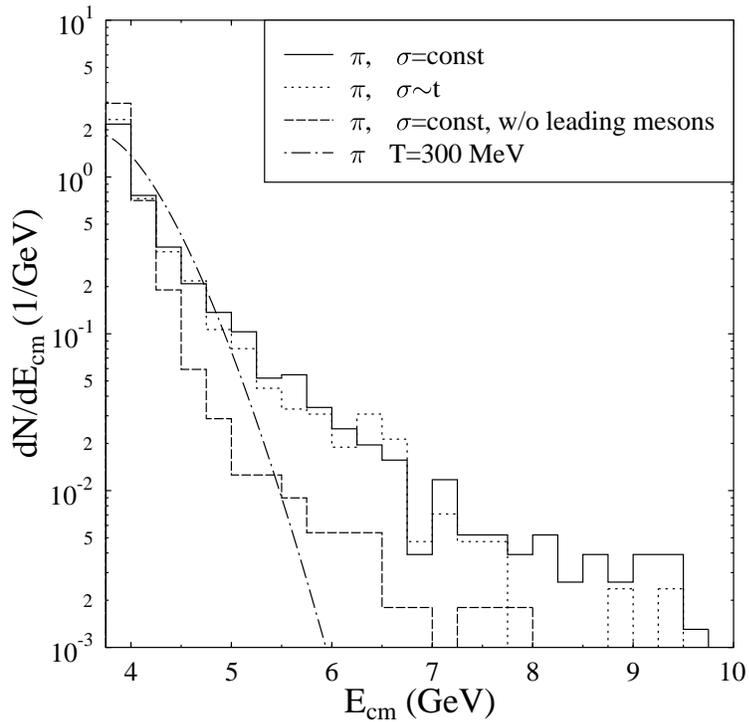,width=12cm}}
\caption{Time integrated $J/\psi\,\pi$ collision spectrum
in central Pb(160~GeV)+Pb reactions according to the UrQMD calculation
(normalized to one collision) in comparison to a thermal hadron gas.
Shown are the energy spectra resulting from 
constant dissociation cross sections including
leading meson interactions (full line), excluding leading mesons, 
Scenario~II (dashed line) and for evolving cross sections with leading mesons,
Scenario~III (dotted line). 
The $J/\psi\,\pi$ collision spectrum in an ideal gas of $\pi$'s at
$T=300$~MeV is indicated by the dash-dotted curve.
\label{spec}}
\vspace*{\fill}
\end{figure}

Figure~\ref{spec} shows the time integrated collision spectrum of
$J/\psi\,\pi$ interactions in central Pb(160~GeV)+Pb reactions according to
the UrQMD simulations. The high
energy tails of the spectra calculated within the different model scenarios
reflect the interactions at early times. They correspond to temperatures of
$T \gg 300$~MeV. However, it is also clear from Fig.~\ref{spec}, that for
all presented scenarios, more than 75\% of $J/\psi\,\pi$ 
interactions occur at collision energies below 4.25~GeV.
Thus, the threshold behaviour of all involved dissociation channels will finally
decide, whether comover absorption is effective enough to explain the
suppression of charmonium yields in nuclear reactions.

\section{Conclusions}
\label{conclusion}
We have studied the production and absorption of charmonium states at SPS
energies using a microscopic hadronic model, UrQMD. The model simultaneously
reproduces the final rapidity distributions of produced hadrons and
net-baryons in $pp$ and $AB$ collisions. The simulations also agree with the
measured total transverse energy spectrum for S(200~GeV)+Au (NA35) and
Pb(160~GeV)+Pb (NA49) as well as with the $E_T$ dependence of the Drell-Yan
muon pairs in Pb(160~GeV)+Pb (NA50). At the same time the calculations are
in significant disagreement with the S(200~GeV)+U $E_T$ dependence of dimuon
production (NA38).

Our study shows that nuclear absorption alone cannot yield prefect
agreement with all measured $J/\psi$ cross sections in $pA$ and $AB$ collisions
at SPS energies. 
This is in
accord with the findings of previous studies with semianalytical models.
However, we find that a global fit to the $AB$ systematics with a nuclear
absorption cross section of 9~mb is not yet completely ruled out by the
data taking into account its systematic and statistical errors and its in
part still prelimary status.

The UrQMD model renders better agreement with the $J/\psi$ data if
comover dissociation is included and the charmonium is
produced in color-singlet states which evolve in time. 
In this case, all parameters were fixed by other data before comparison to
the NA50 data.
The $E_T$ dependence of the $J/\psi$/Drell-Yan ratio is also
reproduced in this model. The $\psi'/J/\psi$ ratio agrees with the $pA$
data, while the $\psi'$ suppression in S+U and Pb+Pb is overestimated.
Assuming produced color-octet states with a time independent, universal
cross section and comover absorption of the fully formed $c\bar c$
states, yields equally good agreement with the $J/\psi$ data. 
The overall trend of the 
$\psi'/J/\psi$ systematic for different projectile and target 
combinations is fairly well reproduced, although significant quantitative
deviations from the data are found in nuclear collisions.

The rapidity dependence of the $J/\psi$ survival probability in $pA$
reactions is a promising observable by which the 
contributions of color-singlet and color-octet states to nuclear 
absorption processes can be disentangled.

The effectiveness of charmonium dissociation by interactions with
secondaries is apparently crucial for the understanding of the experimental data in
hadronic scenarios. The microscopic simulation shows that the
charmonium-comover interactions are not close to an equilibrium situation.
Heavy meson resonances contribute to a large fraction of all
charmonium-comover interactions. The calculation of energy dependent
cross sections for a large number of possible hadronic dissociation channels
is a prerequisite for a quantitative understanding of comover
absorption. 

\begin{table}
\caption{
Comparison of $J/\psi$ survival probabilities for minimum bias Pb+Pb collisions at
different energies for constant and evolving dissociation cross sections, 
{\it i.e.} Scenario~II and Scenario~III, respectively.}\label{table}
\begin{tabular}{|c|c|c|c|c|c|c|}  
\hline
$E_{\rm lab}$ & $S_B^{II}$ & $S_M^{II}$ & $S_{tot}^{II}$ & $S_B^{III}$ & $S_M^{III}$ & $S_{tot}^{III}$ \\ \hline
160~GeV  & 0.51 & 0.61 & 0.31  & 0.57 & 0.49 & 0.33\\
80~GeV   & 0.51 & 0.64 & 0.33  & 0.50 & 0.51 & 0.32
\end{tabular}
\end{table}

\begin{table}
\caption{
Contributions of different meson species to the total 
$J/\psi$ comover absorption in central Pb+Pb collisions 
at 160~GeV according to the UrQMD simulation with Scenario~II
and III.
The percentage of comover absorption due to the most dominant channels 
are shown.}\label{table2}
\begin{tabular}{|c|c|c|}  
\hline
Species    & Scenario~II & Scenario~III \\ \hline
$\pi$      &$ 19.7 \pm 0.3$ &$ 18.6 \pm 0.3 $ \\
$\rho$     &$ 20.8 \pm 0.4$ &$ 18.6 \pm 0.3 $  \\
$K$        &$ 11.6 \pm 0.3$ &$ 11.1 \pm 0.3 $  \\
$K^*(892)$ &$  8.9 \pm 0.3$ &$  8.1 \pm 0.2 $ \\ 
$\eta$     &$  5.4 \pm 0.2$ &$  6.4 \pm 0.2 $  \\
$\omega$   &$  8.4 \pm 0.3$ &$  6.3 \pm 0.2 $  \\
$a_2(1320)$&$ 3.1 \pm 0.1 $ &$  4.4 \pm 0.2 $ \\
$a_1(1260)$&$ 2.6 \pm 0.1 $ &$  4.2 \pm 0.2 $ \\
$b_1(1235)$&$ 3.6 \pm 0.2 $ &$  3.8 \pm 0.2 $ \\
$a_0(980)$ &$  3.0 \pm 0.1$ &$  2.9 \pm 0.1 $ \\ \hline
Sum:       &$ 87.1 \pm 0.8$ &$ 84.4 \pm 0.7 $ 
\end{tabular}
\end{table}

%\begin{figure}[b]
%\vspace*{\fill}
%\centerline{\psfig{figure=/homeb/spieles/plots/carlos/sigab_stopp.eps,width=12cm}}
%\caption{$J/\psi$-production cross sections times dimuon branching ratio
%in the kinematical domain $0<y_{cm}<1$ and $|\cos\theta_{CS}|<0.5$, 
%and rescaled, if necessary, to $p_{\rm lab}=200$~GeV as a function of
%$AB$. The results of the microscopic
%UrQMD simulation without nuclear stopping and momentum distribution of the
%$J/\psi$ are shown as open circles (Scenario~I). Full circles denote the calculation
%where the nuclear stopping is taken into
%account. A universal and constant absorption
%cross section of $\sigma=7.3$~mb is used.
%\label{stopp}}
%\vspace*{\fill}
%\end{figure}
%

%\begin{figure}[b]
%\vspace*{\fill}
%\centerline{\psfig{figure=/homeb/spieles/plots/rapidity/s+u_kahan.eps,width=9cm}
%\hspace*{-1cm}\psfig{figure=/homeb/spieles/plots/rapidity/s+u_evolv.eps,width=9cm}}
%\caption{$J/\psi$-survival probability in S(200~GeV)+U reactions as a
%function of rapidity. Shown are the calculations within Scenario~II (left) and 
%Scenario~III (right).
%\label{rapidity_su}}
%\vspace*{\fill}
%\end{figure}

\end{document}